\documentclass[journal]{IEEEtran}
\usepackage{amsmath}
\usepackage{bm}
\usepackage{amsthm}
\usepackage{graphicx}
\usepackage{float}
\usepackage{subfigure}
\usepackage{cite}
\usepackage{enumerate}
\usepackage{mathrsfs}
\usepackage{multirow}
\usepackage{amsfonts}
\usepackage{makecell}
\usepackage{url}
\usepackage{array}
\usepackage{CJK}
\usepackage{indentfirst}
\usepackage{algorithm}
\usepackage{algpseudocode}
\usepackage{color, soul}
\usepackage{indentfirst} 
\usepackage{colortbl}
\usepackage{threeparttable}
\usepackage{microtype}
\usepackage{xcolor}
\usepackage{diagbox}

\begin{document}
\title{Reconfigurable Holographic Surface:\\ A New Paradigm for Ultra-Massive MIMO}

\author{
	{Boya Di}, \IEEEmembership{Member, IEEE},
	{Hongliang Zhang}, \IEEEmembership{Member, IEEE},
	{Rui Zhang}, \IEEEmembership{Fellow, IEEE},\\
	{Zhu Han}, \IEEEmembership{Fellow, IEEE},
	{and Lingyang Song}, \IEEEmembership{Fellow, IEEE}.
	
	\thanks{B. Di and H. Zhang are with School of Electronics, Peking University, China.}
	\thanks{L. Song is with School of Electronics, Peking University, China, and also with School of Electronic and Computer Engineering, Peking University Shenzhen Graduate School, China.}
	\thanks{R. Zhang is with School of Science and Engineering, Shenzhen Research Institute of Big Data, The Chinese University of Hong Kong, Shenzhen, Guangdong, China, and also with the Department of Electrical and Computer Engineering, National University of Singapore, Singapore.}
	\thanks{H. Zhu is with the Electrical and Computer Engineering Department, University of Houston, USA.}
}

\maketitle
\vspace{-5mm}
\begin{abstract}
Evolving from massive multiple-input multiple-output (MIMO) in current 5G communications, ultra-massive MIMO emerges as a seminal technology for fulfilling more stringent requirements of future 6G communications. However, widely-utilized phased arrays relying on active components make the implementation of ultra-massive MIMO in practice increasingly prohibitive from both cost and power consumption perspectives. In contrast, the development of reconfigurable holographic surface (RHS) provides a new paradigm to solve the above issue without the need of costly hardware components. By leveraging the holographic principle, the RHS serves as an ultra-thin and lightweight surface antenna integrated with the transceiver, which is a promising alternative to phased arrays for realizing ultra-massive MIMO. In this paper, we provide a comprehensive overview of the RHS, especially the RHS-aided communication and sensing. We first describe the basic concepts of RHS, and introduce its working principle and unique practical constraints. Moreover, we show how to utilize the RHS to achieve cost-efficient and high-performance wireless communication and sensing, and introduce the key technologies. In particular, we present the implementation of RHS with a wireless communication prototype, and report the experimental measurement results based on it. Finally, we outline some open challenges and potential future directions in this area.
\end{abstract}


\section{Introduction}

\subsection{Motivation}
Driven by the rapid growth of mobile services, the upcoming sixth generation (6G) wireless networks aim to provide more seamless connectivity and high-speed data transmission \cite{SOBM-2020,KWYJY-2019}. Evolving from massive multiple-input multiple-output (MIMO) used in 5G communications, ultra-massive MIMO has emerged as a key technology to meet the more demanding requirements of 6G. By utilizing the spatial gains with more antennas, ultra-massive MIMO can significantly improve the spectral efficiency and data transmission rate \cite{ZJHWBD-2024}. Therefore, ultra-massive MIMO is regarded as a key enabling technique for 6G, attracting significant interest from both academic researchers and industry professionals.

However, existing MIMO technologies primarily relay on phased arrays, which are facing inherent limitations that hinder them meeting the key performance indicators (KPIs) of 6G, in particular regarding low cost and high energy efficiency. To be specific, phased arrays require a large number of phase shifters and power amplifiers to realize complex phase-shifting circuits, leading to substantial hardware costs and energy consumption, which make it prohibitive to realize ultra-massive MIMO \cite{IHAAKEI-2018}. As a result, the development of innovative technologies are necessary for 6G to support the implementation of ultra-massive MIMO \cite{SHBYMZHL-2022,SJTN-2021,LWR-2024,XQR-2024}.

In contrast to phased arrays, the recent development of metamaterial-based antennas, referred to as \emph{reconfigurable holographic surface} (RHS), has provided a new solution to address the above issues \cite{RBHYL-2022,RBHDZHL-2021}. To be specific, the RHS consists of an internal feed and a large number of meta-material radiation elements. The feed, typically located at the edge or embedded at the bottom of the meta-surface, injects electromagnetic (EM) waves into the meta-surface. By dynamically tuning the EM properties of the radiation elements through simple diodes instead of expensive hardware equipped with phased arrays, the radiated wave from each element can be shaped with low cost and energy consumption. As a result, the RHS is regarded as an effective method to realize the implementation of ultra-massive MIMO. 

Another type of meta-material surface, which is known as reconfigurable intelligent surface~(RIS) or intelligent reflecting surface~(IRS), is extensively studied in the literature \cite{MAMMCJS-2020,QSBCR-2021}. The RIS/IRS consists of an external feed and a large number of meta-material reflective elements. Its working principle involves an EM wave emitted by the feed impinging on each meta-material reflective element, and each element controls the reflection response by adjusting the state of tunable devices, allowing the waves reflected by each element with different phases to coherently superimpose and generate an directional beam \cite{YBN-2022,QR-2019,BHLYZH-2020}. However, due to its external feed and reflective characteristics, the RIS/IRS is generally used as a passive relay in wireless communications, while the RHS is integrated with the transceiver serving as a light-weight and ultra-thin metamaterial antenna.

\begin{figure*}[t]
	\centering
	\includegraphics[width=0.8\textwidth]{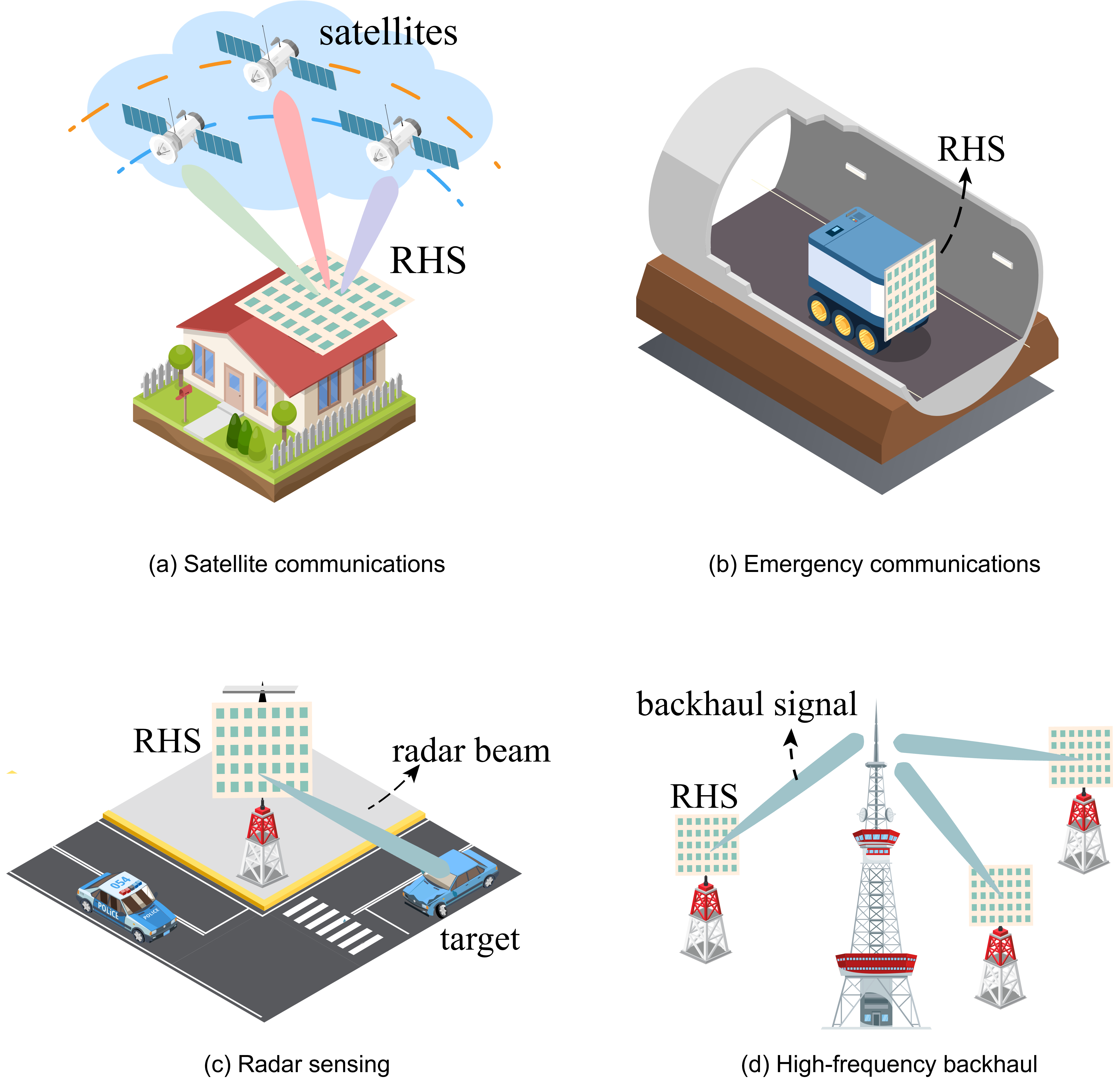}
	\caption{Possible use cases of the RHS.}
	\label{Sec1-Fig:use}
\end{figure*}

\subsection{Use Cases}
Working as a cost-efficient and energy-efficient solution for ultra-massive MIMO, the RHS is in particular suitable for the scenarios with a limited energy supply or payload to carry a bulky antenna array but an aim to achieve higher data rates. As listed in Fig. \ref{Sec1-Fig:use}, the possible use cases of the RHS include:
\begin{itemize}
\item \emph{Satellite communications:} On one hand, the RHS is a light-weight antenna array, which can replace the bulky phased array on satellites to provide high data rates \cite{XYLJ-2025}. On the other hand, the satellite-ground communications suffer severe path loss, the RHS can provide a higher antenna gain to compensate the propagation loss \cite{RBHHL-2022,XRBHL-2023}.

\item \emph{Emergency communications:} As the RHS has an internal feed, it is ultra-thin and lightweight, which makes it promising to be deployed at narrow spaces such as tunnels and mines to provide high data rates \cite{ASDS-2013}.

\item \emph{Radar sensing:} Phased arrays use the phase shifters to provide the electrical beam steering capability for radar detection and the detection resolution is dependent on the number of antennas~\cite{SG-2020}. Considering high energy consumption and hardware cost, the size of phased arrays is limited. The RHS with higher aperture can be a promising solution for the radar sensing application to provide higher sensing resolution.

\item \emph{High-frequency band backhaul:} Communications on high-frequency bands, such as millimeter waves and terahertz bands, can provide wider bandwidth to improve the capacity, which can be a solution for realizing high-speed wireless backhaul, but also are paired with ultra-massive MIMO to provide high direction gains for compensating the severe path loss \cite{MSYLYMGEKCA-2017,ICS-2018}. However, ultra-massive MIMO over the high-frequency bands is more costly as the cost of phase shifters increases with the signal frequency. As the RHS can achieve the beamforming capability without phase shifter, high-frequency band backhual can be a promising use case for the RHS.
\end{itemize}

\subsection{Objective and Organization}
In the literature, some existing articles \cite{QBCLKXWBHEMZR-2024,YRQJHH-2019,HB-2022,MAMMCJS-2020,QSBCR-2021,MHLKZG-2020,YXXTJMN-2021} have provided comprehensive tutorials or surveys of the state-of-the-art research work on metasurface-aided wireless communications, but they focus on the scenario where the surfaces serve as passive relays. In contrast, this tutorial aims to utilize the surface as a cost-efficient solution to achieve the ultra-massive MIMO. Although some tutorials \cite{TPRCGLZMHC-2024,CSGACRMM-2020} have discussed a similar concept of ``holographic MIMO" to the RHS, they did not address the important issues of how to implement such wireless systems and how to design them subject to the practical hardware implementation constraints.  

Specifically, this tutorial provides a comprehensive overview of the latest development and advancement of the RHS. Its main contributions are as follows.
\begin{itemize}
\item First, we present the basic working principle of the RHS with a particular emphasis on explaining its relationship to the concept of \emph{holographic} surface. Also, we highlight the new leakage power constraint introduced by the RHS.

\item Next, we present several case studies in RHS-aided wireless communications and sensing to show its advantages in cost-effectiveness, and the associated new design issues and solutions. 

\item Furthermore, we discuss how to implement the RHS with the illustration of two RHS-aided prototypes including a wireless communication platform and an integrated sensing and communication (ISAC) platform. Based on these prototypes, we show some measurement results to validate the actual performance gain obtained by the RHS in practical systems. 

\item Finally, we outline the major research challenges in RHS-aided wireless networks aiming to stimulate future studies in this area. 
\end{itemize}

\begin{figure}[t]
	\centering
	\includegraphics[width=0.4\textwidth]{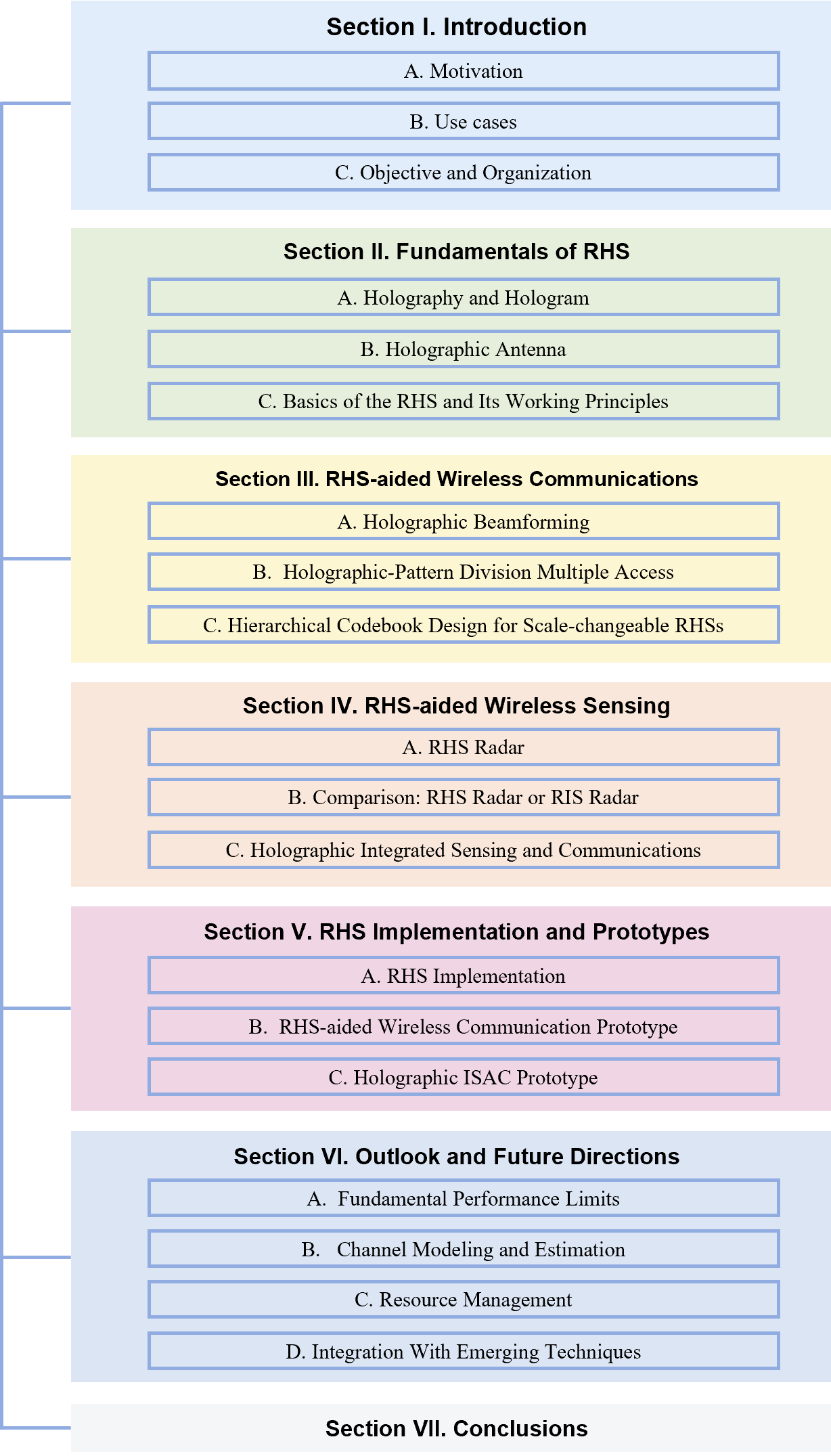}
	\caption{Organization of this paper.}
	\label{Sec1-Fig:organization}
\end{figure}

The rest of paper is organized as shown in Fig. \ref{Sec1-Fig:organization}. Section \ref{Sec:2} introduces the fundamentals of the RHS, including its basic concept and working principle. In Sections \ref{Sec:3} and \ref{Sec:4}, we present the key techniques for the RHS-aided  wireless communications and sensing, respectively. In Section \ref{Sec:5}, we address the implementation of the RHS, and show RHS-aided wireless communication and ISAC prototypes with experimental results. In Section \ref{Sec:6}, we discuss other relevant topics on the RHS to broaden its scope. Finally, we conclude this paper in Section \ref{Sec:7}.

\section{Fundamentals of RHS}
\label{Sec:2}
RHS is a type of holographic antennas which applies the hologram to the antenna design. In the following, we will introduce the basics of RHS, starting from the holography and hologram to the working principle of RHS.

\subsection{Holography and Hologram}
\begin{figure*}[t]
	\centering
	\includegraphics[width=0.8\textwidth]{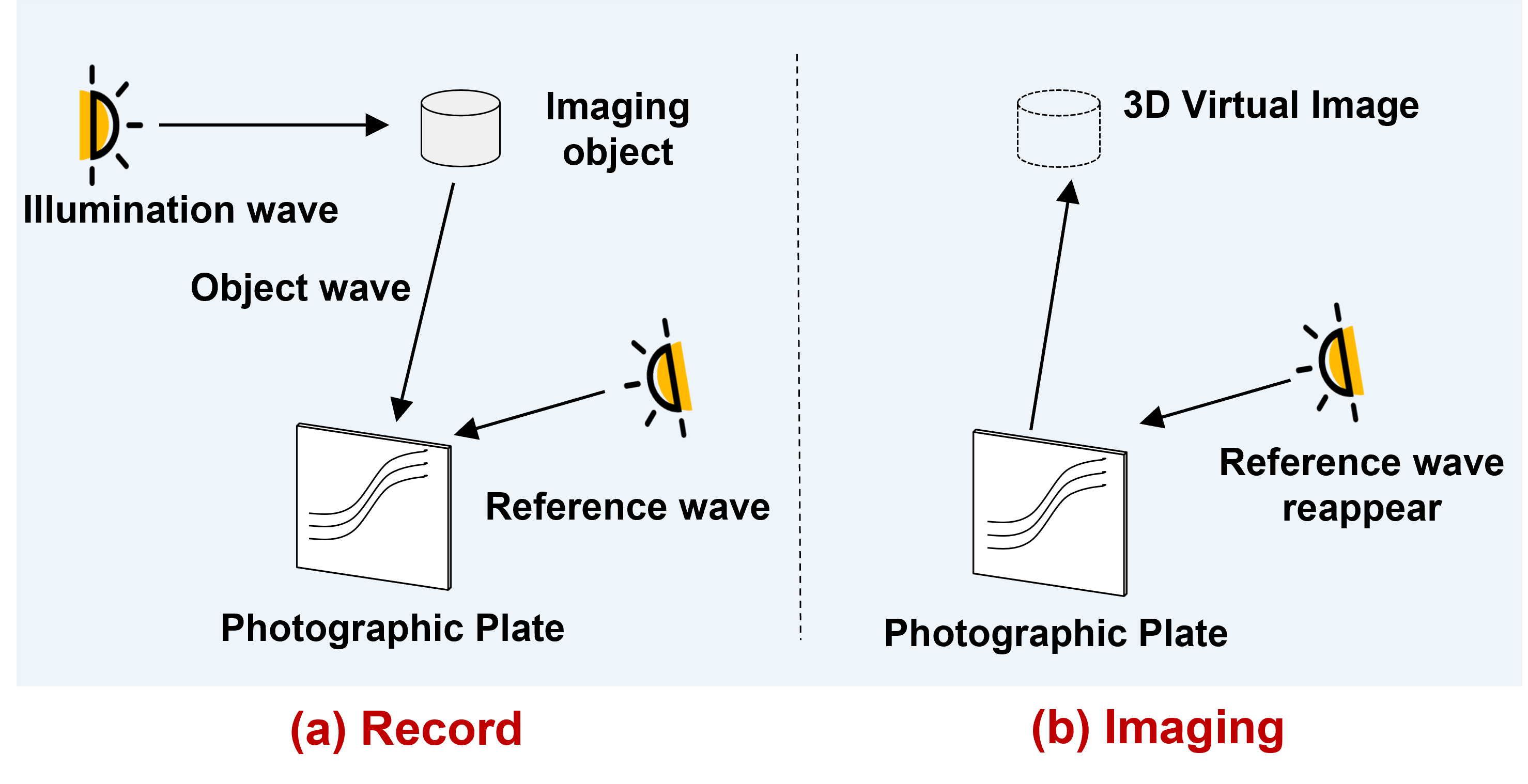}
	\caption{Procedure of the Holography.}
	\label{Sec1-Fig:Hologram}
\end{figure*}

Holography is a technique that enables a wavefront to be recorded and later reconstructed~\cite{P-1996}. It is best known as a method of generating three-dimensional images. A hologram is a recording of any type of wavefront in the form of an interference pattern in the holography, which is also known as \emph{holographic pattern}\footnote{In the following of this paper, we will use the terms of \emph{hologram} and \emph{holographic pattern} interchangeably.}. As shown in Fig. \ref{Sec1-Fig:Hologram}, it takes two steps to generate a virtual images through a hologram.
\begin{itemize}
	\item \textbf{Step 1 Hologram Recording:} We have two sources to generate two coherent light beams. One beam will point to the object. Then, the object will reflect the illuminated wave and generate an object wave. Another beam is called the reference wave. The reference wave interferes with the object wave. The hologram is generated and recorded by a photographic plate.
	
	\item \textbf{Step 2 Imaging Through a Hologram:} We remove the object and the sources. Then, we use the reference wave to illuminate the hologram recorded in the photographic plate, a 3D virtual image will be reconstructed at the same place where the object was placed. 
\end{itemize}

Mathematically speaking, let $\Psi_{obj}$ be the object wave and $\Psi_{ref}$ be the reference wave, respectively. In Step 1, the hologram generated by the interference of object and reference waves can be expressed as $\Psi_{int} = \Psi_{obj}\Psi_{ref}^{*}$. In Step 2, when the reference wave illuminates the hologram, we have $\Psi_{int}\Psi_{ref} \propto \Psi_{obj}\|\Psi_{ref}\|^2$. As the reference wave could be a constant-modulus wave, we can extract the object wave in Step 2, and that is why we can reconstruct the virtual image.

\subsection{Holographic Antenna}
Applying the same idea of holography to the antenna design in the micro-wave bands, a leaky-wave antenna called holographic antenna is proposed \cite{BJJJD-2010}. Here, the object is a beam at a certain direction. The holographic antenna is used to record the micro-wave holographic pattern instead of a photographic plate which records the optical holographic pattern \cite{OD-2017}. In what follows, we will first introduce the hardware structure of the holographic antenna.

\begin{figure}[t]
	\centering
	\includegraphics[width=0.45\textwidth]{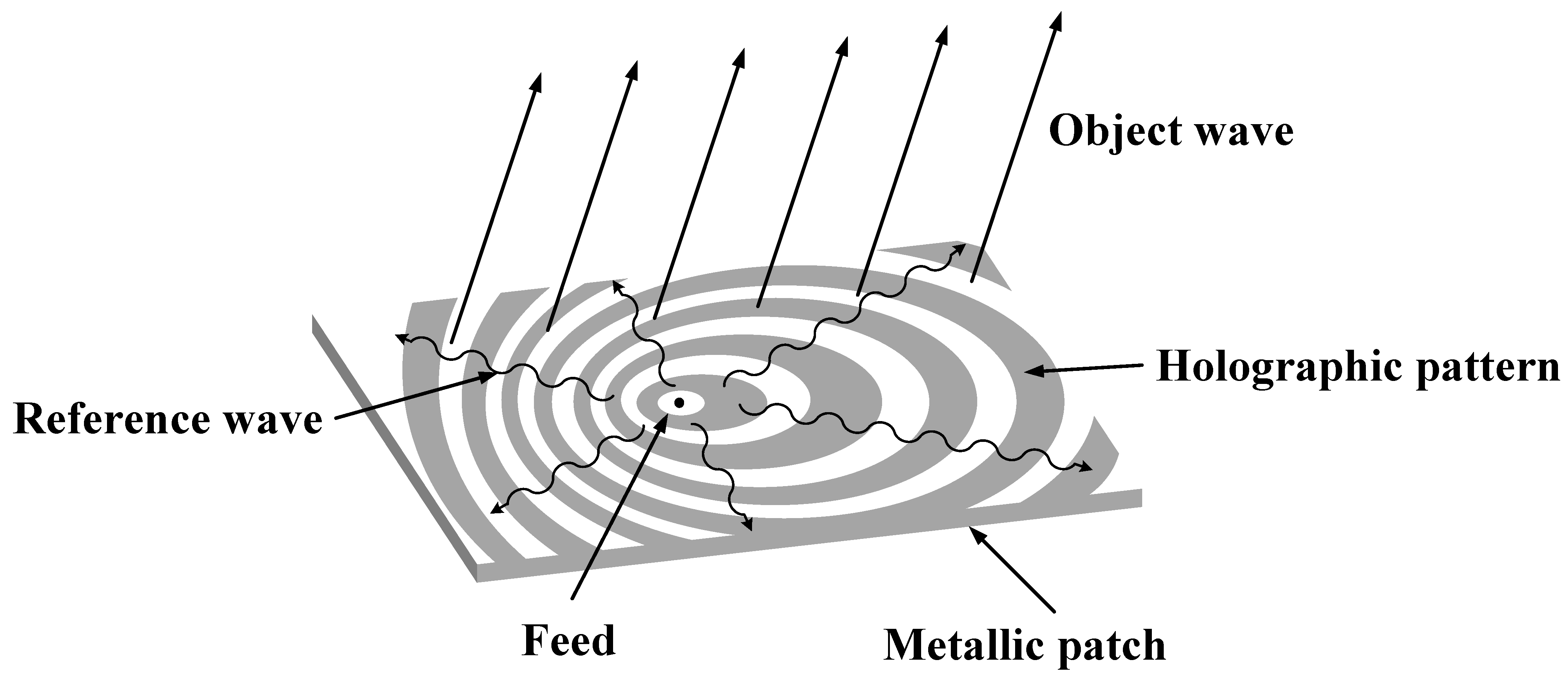}
	\caption{An illustration of the holographic antenna.}
	\label{Sec1-Fig:Holograhpic antenna}
\end{figure}

As shown in Fig. \ref{Sec1-Fig:Holograhpic antenna}, the holographic antenna is a surface which composes of numerous metallic patches. A feed is embedded in the surface. Signals transmitted from the feed, which is carried by the reference wave $\Psi_{ref}$, propagates along the surface. Each metallic patch has its own conditions~(e.g., radiation co-efficient), and all of these patches will contribute to the holographic pattern $\Psi_{inf}$. When the reference wave goes through each patch, a part of the signals will be leaked to the free space and its leakage amount is related to the condition of each patch. Finally, the holographic antenna generates directional beams, i.e., the radiation pattern of the object wave $\Psi_{obj}$.

Likewise, the working procedure of holographic antennas consists of two steps:
\begin{itemize}
	\item \textbf{Holographic Pattern Recording:} The object wave is the beams from desired directions while the signals carrying the information are transmitted from the feed as the reference wave. The holographic pattern is recorded via metallic patch placement to depict the interference between reference and object waves.
	
	\item \textbf{Holographic Beamforming:} Keep the reference wave to interfere with the holographic pattern to recover the directional beams.
\end{itemize}
In this way, the holography is utilized in the antenna design to achieve the function of beamforming. The holographic antennas has the following advantages: 1) \emph{Large scale}: The antenna is composed of numerous metallic patches. Such a structure makes a large-scale antenna array easy to be implemented; 2) \emph{Low cost}: A printed circuit board (PCB) level technique is sufficient for the antenna manufacture; 3) \emph{Lowe power consumption}: There is no complicated phase shifting circuits for beamforming.
 
\subsection{Basics of the RHS and Its Working Principles}
Although the holographic antenna has shown some benefits in wireless communications, it also has a limitation: the placement of the metallic patches is fixed, which makes the beam direction generated by the holographic antenna array unchanged. This makes traditional holographic antenna array inapplicable for mobile scenarios. To address this issue, the concept of RHS is proposed~\cite{RBHDZHL-2021}, which apply reconfigurable metasurface techniques to existing holographic antenna designs. 

\subsubsection{Basics of the RHS}
Metasurface is a thin surface which composes of sub-wavelength elements\cite{MHLKZG-2020}. Controllable components, e.g., PIN diodes, are embedded on each element. With these controllable components, the EM response of each antenna element is adjustable, which makes it possible to generate beams toward different directions with the RHS.

\begin{figure}[t]
	\centering
	\includegraphics[width=0.4\textwidth]{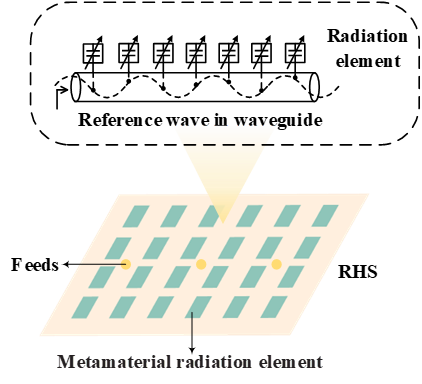}
	\caption{An illustration of the RHS.}
	\label{Sec1-Fig:RHS}
\end{figure}

As shown in Fig. \ref{Sec1-Fig:RHS}, the RHS consists of the following three components \cite{RBHDZHL-2021}:
\begin{itemize}
	\item \textbf{Feed}: The feeds are embedded in the bottom layer of the RHS, which connects to an RF chain, to generate the reference wave, propagating along the surface. Such a structure significantly reduces the thickness of the RHS compared to traditional antenna arrays, where the feeds are outside the surface.
	
	\item \textbf{Waveguide}: The waveguide is the propagation medium of the reference wave. In the RHS, the reference wave is injected into the waveguide which guides the reference wave to propagate on it.
	
	\item\textbf{Metamaterial radiation element}: The metamaterial radiation element is made of artificial material with supernormal EM properties or structures whose responses can be intelligently controlled. Each element is excited by the reference wave such that the radiation characteristic is dictated by the EM responses of each element. Here, the configuration of the RHS is also known as \emph{holographic pattern} which records the directions of desired beams.

\end{itemize}

\subsubsection{Working Principles}
\label{sec:principle}
With the above components, the RHS can configure the direction of generated beams following the idea of holographic antennas. Mathematically speaking, at the $(n_y,n_z)$-th radiation element, the reference wave generated by feed $k$  and the wave propagating in free space with a direction of $(\theta_0, \varphi_0)$, which is also referred to as the objective wave, can be given by \cite{MSN-2016}
\begin{equation}\label{ref}
	\Psi_{ref}(\mathbf{r}_{n_y,n_z}^k)=\exp(-j\mathbf{k}_s\cdot\mathbf{r}_{n_y,n_z}^k),
\end{equation}
\begin{equation}
	\Psi_{obj}(\mathbf{r}_{n_y,n_z}, \theta_0, \varphi_0)=\exp(-j\mathbf{k}_f\cdot\mathbf{r}_{n_y,n_z}),
\end{equation}
where $\mathbf{k}_s$ is the propagation vector of the reference wave, $\mathbf{r}_{n_y, n_z}^k$ is the distance vector from the feed $k$ to the $(n_y, n_z)$-th radiation element, and  $\mathbf{k}_f$ is the desired directional propagation vector in free space.
The interference between the reference wave and the object wave is defined as
\begin{equation}\label{intf}
	\Psi_{intf}(\mathbf{r}_{n_y,n_z}^k, \theta_0, \varphi_0)=\Psi_{obj}(\mathbf{r}_{n_y,n_z}, \theta_0, \varphi_0)\Psi_{ref}^{\ast}(\mathbf{r}_{n_y,n_z}^k).
\end{equation}
The information contained in the holographic pattern $\Psi_{intf}$ is supposed to be recorded by the radiation elements. When the holographic pattern is excited by the reference wave, we have
$\Psi_{intf}\Psi_{ref}\propto \Psi_{obj}|\Psi_{ref}|^2$.
Therefore, the wave propagating in the direction $(\theta_0, \varphi_0)$ is generated.

To construct the holographic pattern given in (\ref{intf}), each radiation element controls the radiation amplitude of the reference wave. It should be noted that the phase of the reference wave at each element is fixed, which should be considered in the antenna design. To be specific, the phase of the reference wave is determined by the relative position of each element on the surface as given in (\ref{ref}). Intuitively speaking, the radiation elements whose phases are aligned with the object wave will be controlled to radiate more (higher amplitude), while those are out of phase will be tuned to radiate less (lower amplitude) in order to concentrate the energy in a target direction \cite{RBHYL-2021}.

The real part of the interference (i.e., $\text{Re}[\Psi_{intf}]$), i.e., cosine value of the phase difference between reference and object waves, decreases as the phase difference grows, which exactly meets the above amplitude control requirements. To capture this feature, we use $\text{Re}[\Psi_{intf}]$ to represent the radiation amplitude of each element. To avoid generating a negative value, $\text{Re}[\Psi_{intf}]$ is normalized to $[0, 1]$. Mathematically speaking, the holographic pattern $\bm{m}$ to generate a beam in the direction $(\theta_0, \varphi_0)$ can be expressed as
\begin{equation}\label{mt}
	\bm{m}(\theta_0, \varphi_0)=\frac{\text{Re}[\Psi_{intf}(\theta_0, \varphi_0)]+1}{2},
\end{equation}
and the radiation amplitude of each element with holographic pattern $\bm{m}(\theta_0, \varphi_0)$ can be written as
\begin{equation}\label{m1}
	m(\mathbf{r}_{n_y,n_z}^k,\theta_0, \varphi_0)=\frac{\text{Re}[\Psi_{intf}(\mathbf{r}_{n_y,n_z}^k, \theta_0, \varphi_0)]+1}{2}.
\end{equation}

\subsubsection{Leakage Power Constraints}
\begin{figure}[t]
	\centering
	\includegraphics[width=0.45\textwidth]{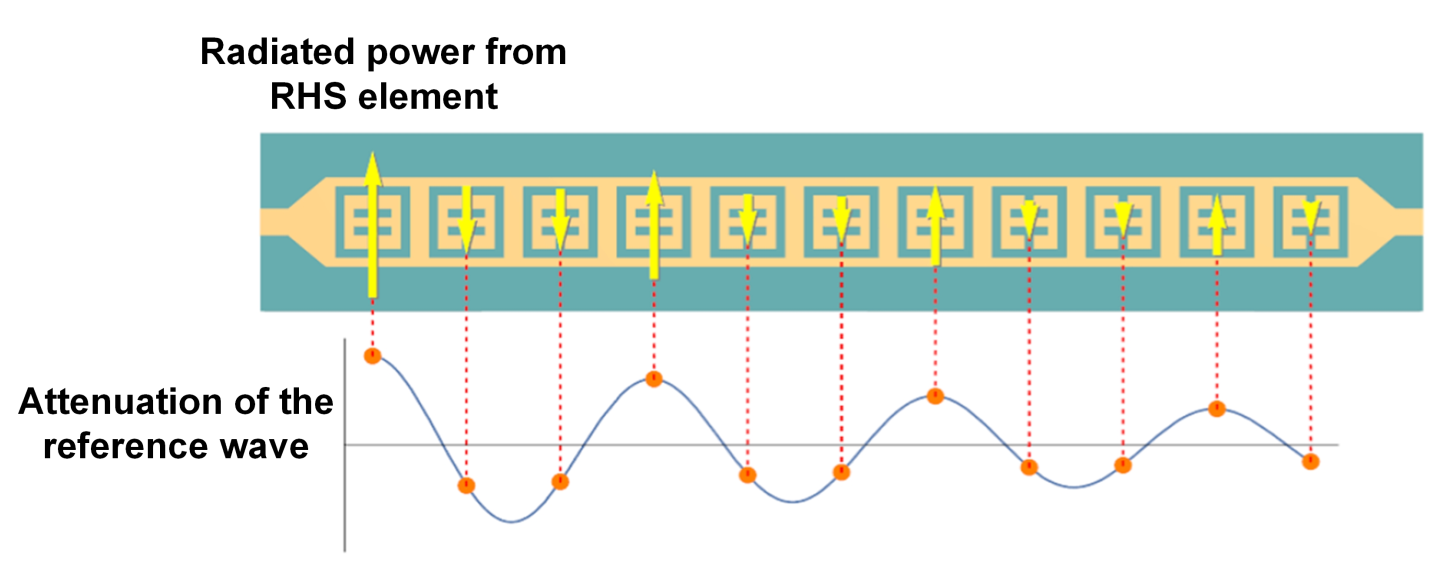}
	\caption{Illustration of the leakage power constraint.}
	\label{Sec1-Fig:leakage}
\end{figure}

According to the RHS's working principle, the RHS also brings an unique constraint: leakage power constraint. To be specific, the RHS is a type of leaky-wave antenna and adopts a series feeding mechanism where the incident reference wave propagates along the waveguide, exciting each element one by one. This process transforms the reference wave into a leaky wave via the elements' slots, facilitating signal emission into free space. Such an operating mechanism makes a gradual attenuation of the reference wave during its propagation process, as shown in Fig. \ref{Sec1-Fig:leakage}. Hence, the RHS obeys a leakage power constraint, i.e., the sum of the radiated power from RHS elements should be no larger than the transmit power from the feed, i.e., 
\begin{equation}\label{leakage_con}
	\sum\limits_{n_y}\sum\limits_{n_z} \eta_{n_y,n_z} \left(m(\mathbf{r}_{n_y,n_z}^k,\theta_0, \varphi_0)\right)^2 \leq P_t,
\end{equation}
where $\eta_{n_y,n_z}$ is the ratio of the power accepted by element $(n_y,n_z)$ to the total transmit power, which is generally less than 1, and $P_t$ is the total transmit power. It should be noted that the radiated power of the current element is related to those of previous elements, which results in a coupling among RHS elements, thereby requiring new methods to design the  holographic beamforming scheme.

\subsubsection{Comparison of the RHS and the RIS/IRS}
After the introduction of the RHS, one may have a question: what is the difference between the RHS and the RIS/IRS. In general, these two types of surfaces are different in the following three aspects, as summarized in Table \ref{difference}.
\begin{itemize}
	\item \textbf{Physical Structure:} As introduced before, the feed (the RF front end) is embedded in the RHS. As a result, the whole antenna (including the surface and the RF front) can be thin. For the RIS/IRS, the RF front end is outside the surface, and therefore, the whole antenna is much more bulky than the RHS.
	
	\item \textbf{Operating Mechanism:} The RHS is a leaky-wave antenna, i.e., the reference wave propagates along the waveguide and a part of the energy was leaked into the space for radiation. For the RIS/IRS, it reflects the incident signals toward the target direction. This also leads to different feeding types. For the RHS, the reference wave is radiated from antenna elements one by one, i.e., serial feeding. In contrast, the antenna elements of the RIS/IRS reflect the signals simultaneously, i.e., parallel seeding.
	
	\item \textbf{Typical Applications:} As the RHS is thin and light-weight, it is used as transmit/receive antennas, and it can be mounted on mobile platforms. Due to the reflection characteristics of the RIS/IRS, it is widely used as passive relays, which can be deployed in the cell edge for coverage extension.
\end{itemize}

\begin{table*}[t]
	\footnotesize
	\centering
	\caption{Difference between the RHS and the RIS/IRS}
	\label{difference}
	\begin{tabular}{m{1cm}|m{4cm}|m{3.5cm}|m{4.5cm}}
		\hline
		\hline
		\textbf{Type} & \textbf{Physical Structure}&\textbf{Operating Mechanism} & \textbf{Typical Applications} \\
		\hline
		RHS & RF front end is \textbf{integrated} into the surface & \makecell[l]{\textcircled{1} Leaky-wave antenna
		\\ \textcircled{2} Serial feeding} & \makecell[l]{\textcircled{1} Transmit/receive antennas
		\\ \textcircled{2} Mounted on mobile platforms}\\
		\hline
		RIS/IRS & RF front end is \textbf{outside} of the surface & \makecell[l]{\textcircled{1} Reflection antenna\\ \textcircled{2} Parallel feeding} & \makecell[l]{\textcircled{1} Passive relays\\ \textcircled{2} Deployed in the cell edge} \\
		\hline
	    \hline
	\end{tabular}
\end{table*}

\section{RHS-aided Wireless Communications}
\label{Sec:3}
In this section, we will show several case studies about how to incorporate the RHS into wireless communication systems. In Section~\ref{Sec:holographic-beamforming}, we will introduce the holographic beamforming scheme to support multi-user wireless communications with an RHS, and design a holographic-pattern division multiple access~(HDMA) solution to reduce the complexity by exploiting the advantage of the amplitude control provided by the RHS in Section~\ref{Sec:HDMA}. Finally, in Section~\ref{Sec:codebook}, we present a codebook-based beamforming scheme to bypass the channel estimation overhead, and show the RHS as a scale-changeable antenna to adaptively change the beamwidth for various scenarios.

\subsection{Holographic Beamforming}
\label{Sec:holographic-beamforming}
\begin{figure}[t]
	\centering
	\includegraphics[width=0.45\textwidth]{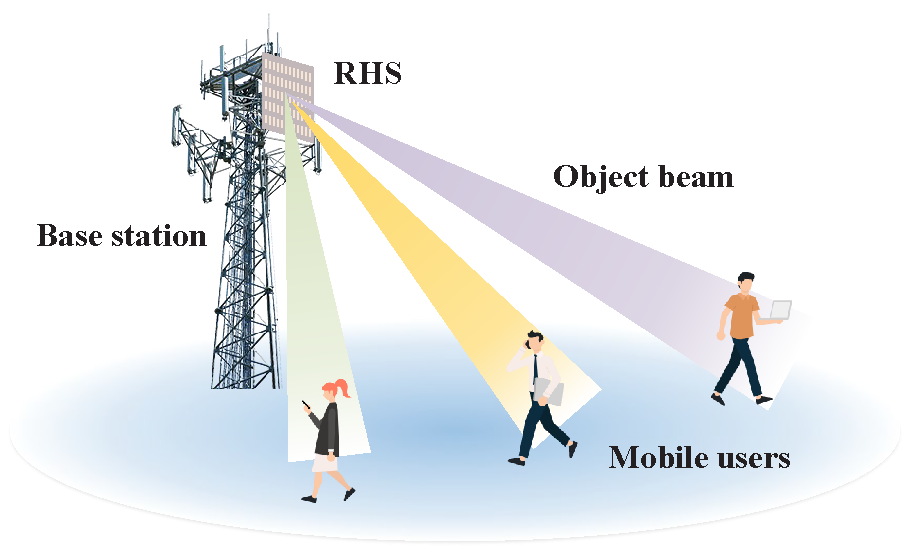}
	\caption{RHS-aided multi-user communication systems.}
	\label{system_model1}
\end{figure}

Consider a downlink multi-user communication system in Fig. \ref{system_model1} as an example, where a BS transmits data streams to $L$ mobile users, each corresponding to a single data stream from the BS. Without loss of generality, it is assumed that each user is equipped with $M$ antennas and a single RF receive chain. The BS is equipped with an RHS to generate beams for users with the beamforming scheme \cite{RBHYL-2022}.
\subsubsection{Hybrid Beamforming Scheme}
\begin{figure}[t]
	\centering
	\includegraphics[width=0.45\textwidth]{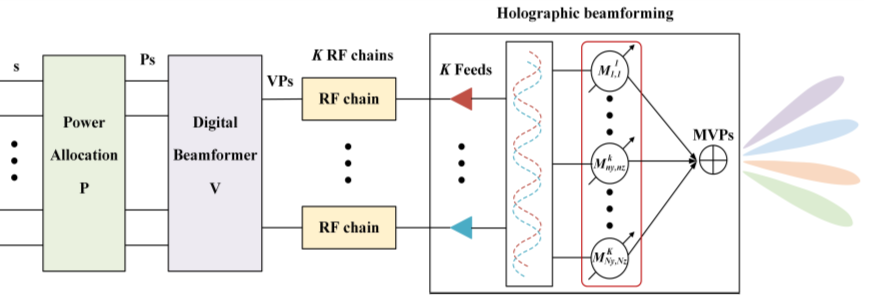}
	\caption{Block diagram of the hybrid beamforming for RHS-aided communication systems.}
	\label{block1}
\end{figure}

Similar to the traditional MIMO enabled by phased arrays, the BS is required to carry out the signal processing at the baseband. As shown in Fig.~\ref{block1}, the BS encodes $L$ data streams via a digital beamformer $\mathbf{V}$, and then up-converts the processed signals to the carrier frequency through RF chains. Each RF chain is connected with a feed of the RHS\footnote{It is assumed that the number of RF chains is no less than the number of data streams\cite{FW-2016}. In other words, the number of feeds should also be no less than the number of active data streams.}. The feed then transforms the up-converted signals into a reference wave propagating on the RHS. The analog beamforming is done at the RHS by controlling the radiation amplitude (holographic pattern) at each element, i.e., $\{m_{n_y,n_z}\}$, to generate desired beams, which is also referred to as \emph{holographic beamformer}. 

\subsubsection{Holographic Beamformer Optimization}
\label{holographic-beamformer}
Denote the intended signal for mobile users as $\bm{s}$, and thus the received signals at mobile user $l$ can be given by
\begin{equation}
	y_l=\bm{W}_l^{H}\bm{H}_l\bm{M}\bm{V}_l\bm{s}_l+\bm{W}_l^{H}\bm{H}_l\bm{M}\sum_{l'\neq l}\bm{V}_{l'}\bm{s}_{l'}+\bm{W}_l^{H}\bm{z}_l,
\end{equation}
where $\bm{M}$ is the holographic beamformer matrix whose element is $\{m_{n_y,n_z}\cdot e^{-j\mathbf{k}_s\cdot\mathbf{r}_{n_y,n_z}^k}\}$ with $e^{-j\mathbf{k}_s\cdot\mathbf{r}_{n_y,n_z}^k}$ being the phase of the reference wave at $(n_y,n_z)$-th element, $\bm{V}_l$ is the $l$-th column of $\bm{V}$, and $\bm{z}_l$ is the additive white Gaussian noise. $\bm{W}_l$ is the RF combiner at mobile user $l$ where $|\bm{W}_l|^2 = 1$, and $\bm{H}_l$ is the matrix of complex channel gains from the RHS to the user $l$. Therefore, the achievable rate of user $l$ can be given by
\begin{equation}\label{Rl}
	R_l=\log_2\left(1+\frac{|\bm{W}_l^{H}\bm{H}_l\bm{M}\bm{V}_l|^2}{J\sigma^2+\sum_{l'\neq l}|\bm{W}_l^{H}\bm{H}_l\bm{M}\bm{V}_{l'}|^2}\right).
\end{equation}

The optimization problem aims to maximize the sum rate of all the users by optimizing digital beamformer $\bm{V}$, holographic beamformer $\bm{M}$, and combiner $\bm{W}$, subjected to the power constraint. Mathematically, the optimization problem can be expressed as
\begin{subequations}\label{max}
	\begin{align}
		\max_{\{\bm{W},\bm{V},\bm{M}\}}&\sum_{l=1}^LR_l\\
		s.t.\quad & |\bm{W}_l|^2 = 1, \\
		& \text{Tr}(\bm{MV}\bm{V}^{H}\bm{M}^{H})\leq P_T, \label{con2}\\
		& 0\leq {m}_{n_y,n_z}\leq 1, \forall n_y,n_z, \label{mc_b}
	\end{align}
\end{subequations}
where $P_T$ is the total transmit power of the BS, and (\ref{con2}) comes from the leakage power constraint of the RHS provided in (\ref{leakage_con}).

To solve this problem, we optimize the digital beamformer, the holographic beamformer, and the combiner one by one. The optimizations of the digital beamformer and the combiner are the same as traditional MIMO systems. For example, the digital beamformer can be a zero-forcing beamformer as a low-complexity but near-optimal solution~\cite{FDBETF-2013}. However, the analog beamformer enabled by the RHS uses amplitude control (holographic beamformer) to replace the phase control with phased arrays, which makes the optimization variables real-valued instead of complex-valued. Due to the existence of the multi-user interference, the problem is still non-convex. Therefore, the basic idea is to recast the data rate into a maximization problem of a concave function based on fractional programming technique~\cite{KW-2016}. After that, the problem can be transformed into a convex problem, which can be solved by existing convex optimization techniques \cite{SL-2004}. As the optimization variables are changed from the complex-value domain to the real-value domain, which significantly reduces the computational complexity. For more details, please refer to \cite{RBHYL-2022}.

\subsubsection{Simulation Results}
\begin{figure}[t]
	\centering
	\includegraphics[width=0.45\textwidth]{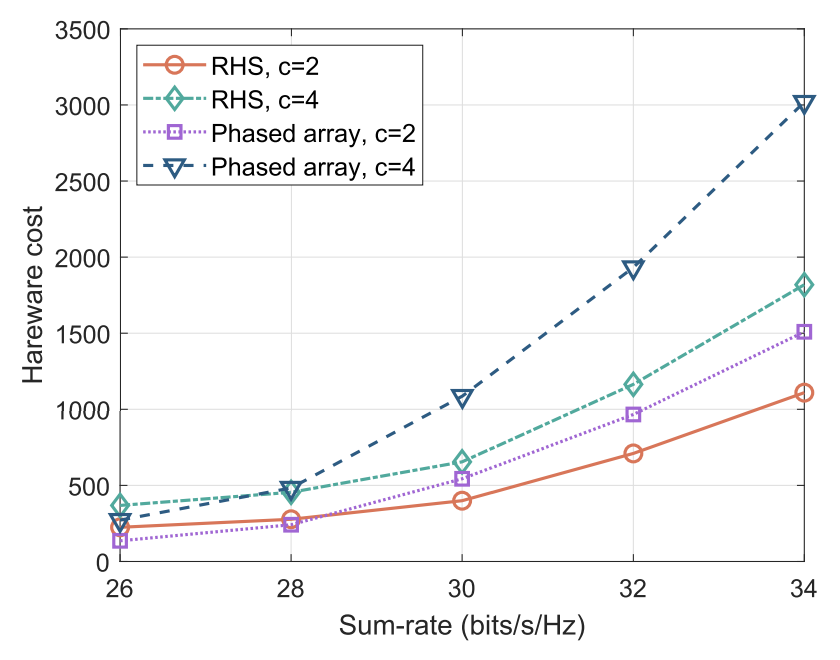}
	\caption{Comparison between the RHS and the phased array in terms of the hardware cost.}
	\label{Sec2:Result1}
\end{figure}
In Fig. \ref{Sec2:Result1}, we compares the hardware cost of the RHS and that of the phased array for a equal sum-rate requirement. The major cost of a phased array is in the Transmit (Tx)/Receive (Rx) modules while the major cost of an RHS comes from the radiation element. For convenience, we define $c$ as the cost ratio of the phased array's Tx/Rx module to the radiation element. From this figure, we can observe that when the cost ratio $c$ and the sum-rate requirement are low, the hardware cost of the phased array is lower than that of the RHS. The main reason is that the phased array can achieve the required sum-rate with fewer elements. However, as the required sum-rate grows, the RHS can achieve the sum-rate at a lower hardware cost. Although the phased array requires fewer elements to achieve the required sum-rate, the reduced number of elements is still insufficient to offset its high hardware cost. The cost efficiency gain provided by the RHS becomes more significant with a higher sum-rate requirement, which meets the future wireless network trend.

\subsection{Holographic-Pattern Division Multiple Access}
\label{Sec:HDMA}
In Section \ref{Sec:holographic-beamforming}, we have introduced the holographic beamforming method for multiple access. However, the number of variables for holographic beamforming equals to the number of RHS elements. In other words, the computational complexity scales up with the growing number of RHS elements for ultra-massive MIMO. To reduce the computational complexity, we leverage the advantage of the amplitude control brought by the RHS to propose a new multiple access scheme, i.e., holographic-pattern division multiple access (HDMA). 

\subsubsection{Principle of HDMA}
The main characteristic of the HDMA is to utilize the superposition of all the holographic pattern to map all transmitted data (carried by the reference waves) to a single-user holographic pattern on the RHS. To be specific, we can obtain the multi-user holographic pattern by a weighted sum of all the single-user holographic patterns rather than the direct holographic beamformer optimization. In this way, it is no need to compute the holographic beamformer frequently in dynamic scenarios, and we can maintain single-user holographic patterns for each user and add them together to reduce the computational complexity. It should be noted that the HDMA method is to replace the holographic beamforming step introduced in Section \ref{holographic-beamformer}, the digital beamformer is still needed to reduce the multi-user interference. The specific principles of HDMA is elaborated as below.

\begin{figure}[t]
	\centering
	\includegraphics[width=0.45\textwidth]{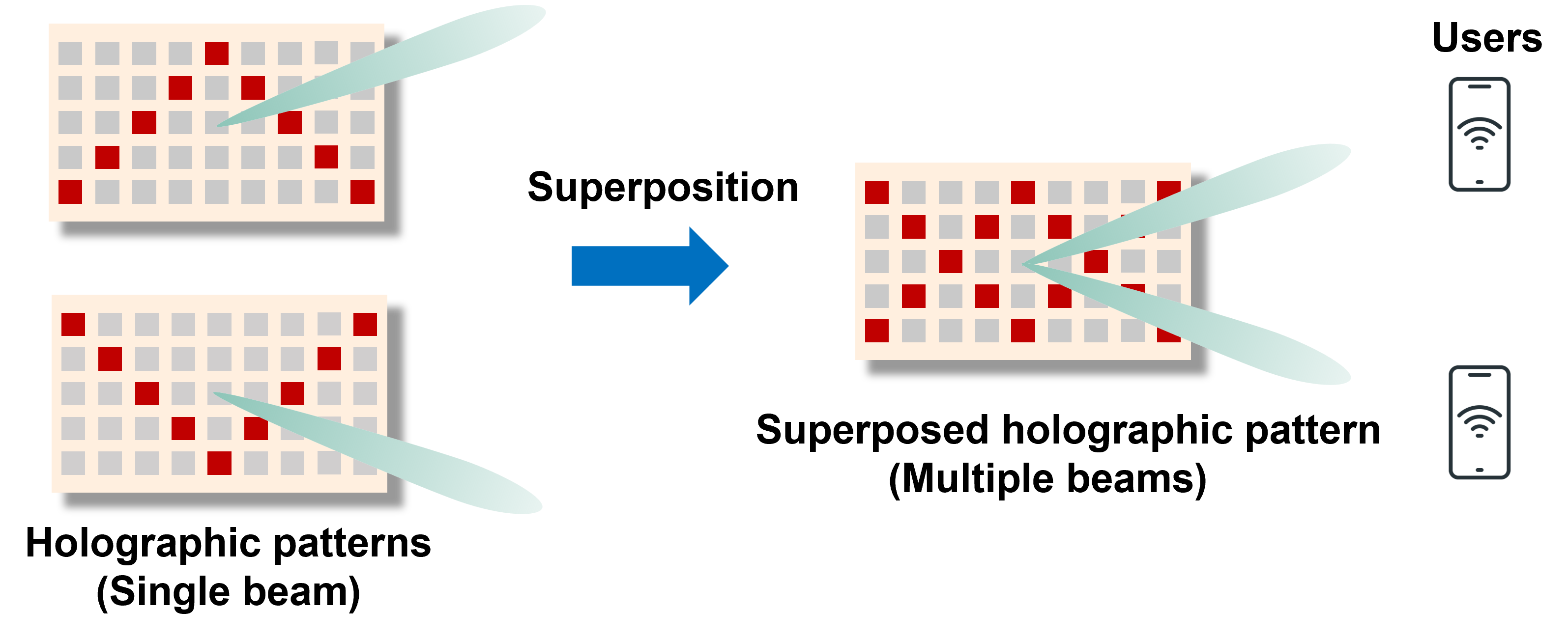}
	\caption{Illustration of the HDMA.}
	\label{HDMA}
\end{figure}

As introduced in Section \ref{sec:principle}, the holographic pattern is generated through the interference between object and reference waves. According to the key idea of the HDMA, a multi-user holographic pattern can be calculated as a weighted summation of single-user holographic patterns each corresponding to a user's beam according to~(\ref{m1}). Mathematically, a $L$-user holographic pattern $\bm{m}$, i.e., the normalized radiation amplitude of each radiation element, can then be given by
\begin{equation}\label{m2}
	m_{n_y, n_z}=\sum_{l=1}^L\sum_{k=1}^Ka_{l,k}m(\mathbf{r}_{n_y, n_z}^k,\theta_l,\varphi_l),
\end{equation}
where $a_{l,k}$ is the weight of the data to user $l$ from feed $k$ satisfying
$\sum_{l=1}^L\sum_{k=1}^Ka_{l,k}=1$. This constraint is set to guarantee that the radiation amplitude of each RHS element satisfies $m_{n_y,n_z} \in [0,1]$.

\textbf{An illustrative example for HDMA:} Fig. \ref{HDMA} shows an two-user example of data mapping according to the HDMA. The feeds of the RHS are assumed to be located close to the origin, where $\mathbf{r}_{n_y, n_z}^k\approx\mathbf{r}_{n_y, n_z}, \forall k$. This is easy to be achieved with a large antenna aperture. Assume that the zenith and azimuth angles of user is $\theta_l$ and $\varphi_l$, respectively. Based on~(\ref{m1}), we have
$m(\mathbf{r}_{n_y, n_z}^1,\theta_l,\varphi_l)=m(\mathbf{r}_{n_y, n_z}^2,\theta_l,\varphi_l)=\cdots=m(\mathbf{r}_{n_y, n_z}^K,\theta_l,\varphi_l)=m(\mathbf{r}_{n_y, n_z},\theta_l,\varphi_l)$,
and thus, the index $k$ can be omitted. The normalized radiation amplitude of each radiation element in the holographic pattern $\bm{m}_l$ corresponding to user $l$ can then be denoted as $\{m(\mathbf{r}_{n_y, n_z},\theta_l,\varphi_l)\}$.
In the HDMA scheme, user 1 and user 2 are multiplexed on a holographic pattern, i.e., the RHS maps the intended signals for these two users onto a multi-user holographic pattern superposed by single-user holographic patterns $\bm{m}_1$ and $\bm{m}_2$, which can be given by
\begin{equation}
	\begin{split}
		&\bm{m}=a_{1}\bm{m}_1+a_{2}\bm{m}_2~\Leftrightarrow \\
		&m_{n_y, n_z}=a_{1}m(\mathbf{r}_{n_y, n_z},\theta_1,\varphi_1)+a_{2}m(\mathbf{r}_{n_y, n_z},\theta_2,\varphi_2).
	\end{split}
\end{equation}

\subsubsection{Holographic Pattern Optimizations}
According to the HDMA principle, a multi-user holographic pattern can be obtained by a weighted sum of multiple single-user holographic patterns. As a result, the multi-beam holographic pattern optimization can be achieved via optimizing the weighting factor $a_{l}$. According to the results in \cite{RBHL-2022}, the optimal weighting factor $a_{l}^{\ast}$ satisfies
\begin{equation}\label{oa}
	\left\{
	\begin{split}
		&a_l^{\ast}=\frac{1}{\beta^{\ast}\ln2}+\sqrt{\frac{1}{(\beta^{\ast}\ln2)^2}-\frac{1}{I_l}},\\
		&\sum_{l=1}^La_l^{\ast}=1,
	\end{split}
	\right.
\end{equation}
where $\beta^{\ast}$ is a normalized factor and $I_l$ is a constant related with the position of user $l$. The complexity of the optimization problem is only related to the number of users while that of the holographic beamforming scheme given in Section \ref{Sec:holographic-beamforming} is proportional to the number of antenna elements. In most use cases, the number of users are much lower than the number of antennas, which indicates that the HDMA scheme can significantly reduce the complexity.

\subsubsection{Simulation Results}

\begin{figure}[t]
	\centering
	\subfigure[]
	{
		\label{Sec2-Fig:capacity}
		\includegraphics[width=0.4\textwidth]{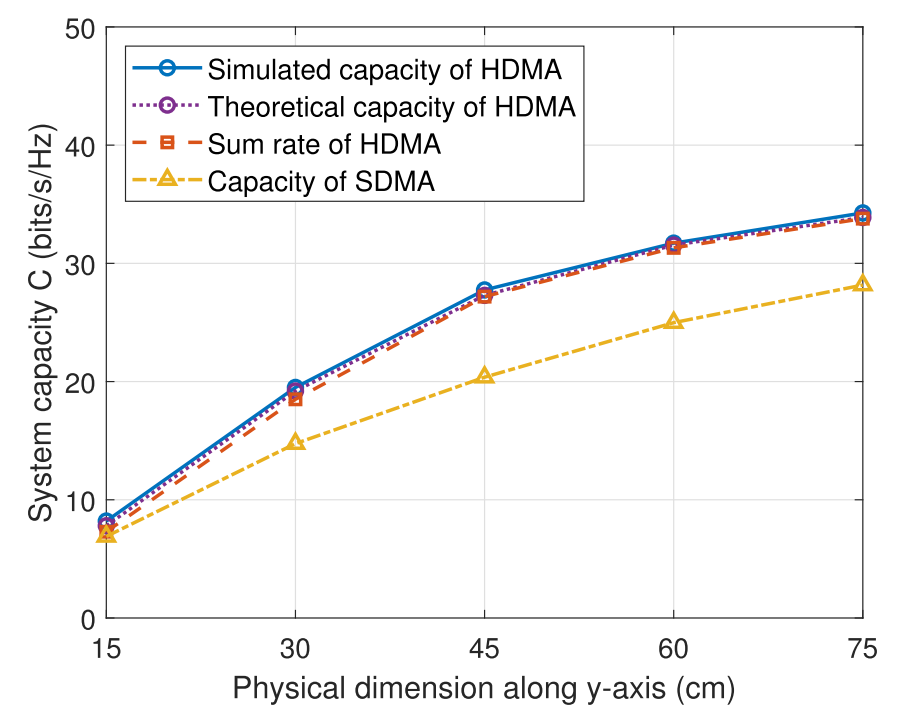}
	}
	\subfigure[]
	{
		\label{Sec2-Fig:cost}
		\includegraphics[width=0.4\textwidth]{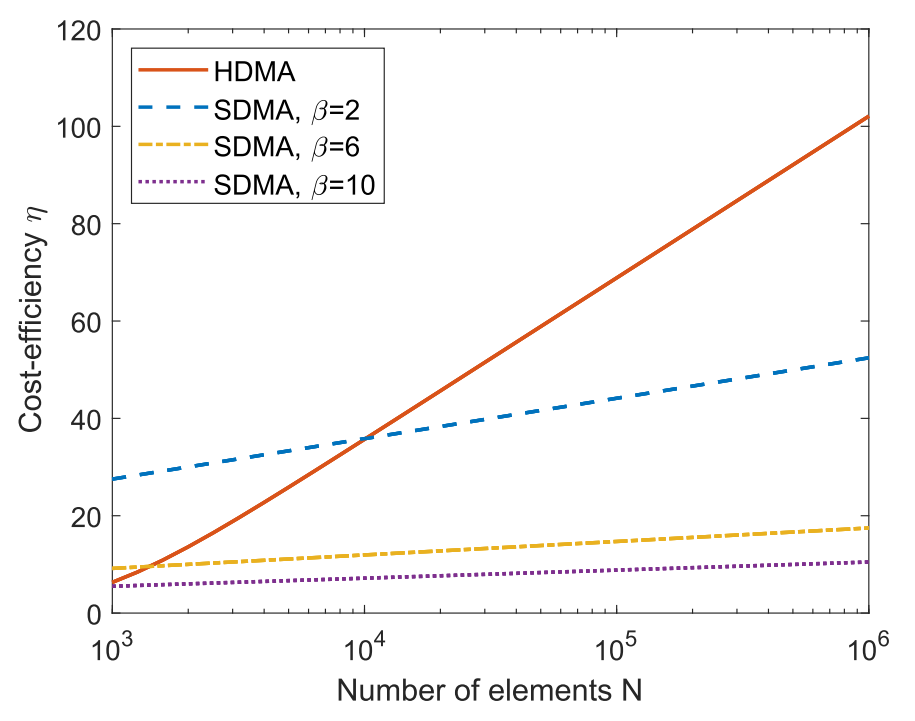}
	}
	\caption{Performance Analysis: (a) System capacity vs. physical dimension; and (b) Cost efficiency vs. number of elements.}
		\vspace{-0.5cm}
	\label{HDMA-Result}
\end{figure}

In Fig. \ref{HDMA-Result}, we compare the performance of the HDMA scheme with that of existing space-division multiple access (SDMA) scheme \cite{DMDPA-2003}. In \ref{Sec2-Fig:capacity}, we depicts the capacity using the HDMA scheme versus the physical dimension of the RHS. We observe that even for a normal-sized RHS, the sum rate of the HDMA scheme can still achieve its capacity. Fig. \ref{Sec2-Fig:cost} shows the cost-efficiency, where the physical dimensions of the RHS and the phased array are equal. Here, the cost-efficiency $\eta$ is defined as the sum rate over the hardware cost. Typically, the hardware cost ratio of phased array to the RHS for each element $\beta$ is $2\sim 10$, since the phased array requires high-priced electronic components such as phase shifters~\cite{Pi-2019}. It can be observed that the cost-efficiency of the HDMA scheme is first lower and then greater than that of the SDMA scheme. This indicates that the superiority of the HDMA scheme in cost savings becomes clearer with a larger size of the RHS and a higher cost ratio $\beta$.


\subsection{Hierarchical Codebook Design for Scale-changeable RHSs}
\label{Sec:codebook}

In previous parts, we have introduced the multiple access schemes for the RHS-aided wireless communication systems when the instantaneous CSI is known. However, as the size of the RHS increases, the CSI estimation costs increase prohibitively, which makes it challenging to obtain the accurate instantaneous CSI for beamforming. The codebook based schemes are potential solutions to address the above issues \cite{HB-2022}. To be specific, a codebook consists of multiple predefined codewords, each corresponding to a beam radiation pattern. By returning the received signal strength, the BS can know the which codeword can lead to a better system performance, for example, a higher data rate, bypassing the channel estimation. 

Moreover, as the aperture of the RHS expands, the corresponding near-field region\footnote{The boundary of near and far fields is typically determined by the Rayleigh distance\cite{HSHTH-2024}, which is positively proportional to the antenna aperture.} expands as well. For the near-field region, the codebook design should be formulated in the angle-distance domain, instead of the pure angular domain for the far-field region~\cite{YZJCXR-2023}, which may bring substantial beam training overhead. To reduce the overhead, hierarchical codebooks with multi-coverage codewords are proposed~\cite{YBHL-2023}. However, it is challenging to generate codewords with different coverage regions but flat gain for a fixed array size. A possible solution is to utilize the RHS as switchable antennas to address this issue~\cite{SBHH-2024}.

\subsubsection{Scale-changeable RHS}
\begin{figure}[t]
	\centering
	\includegraphics[width=0.4\textwidth]{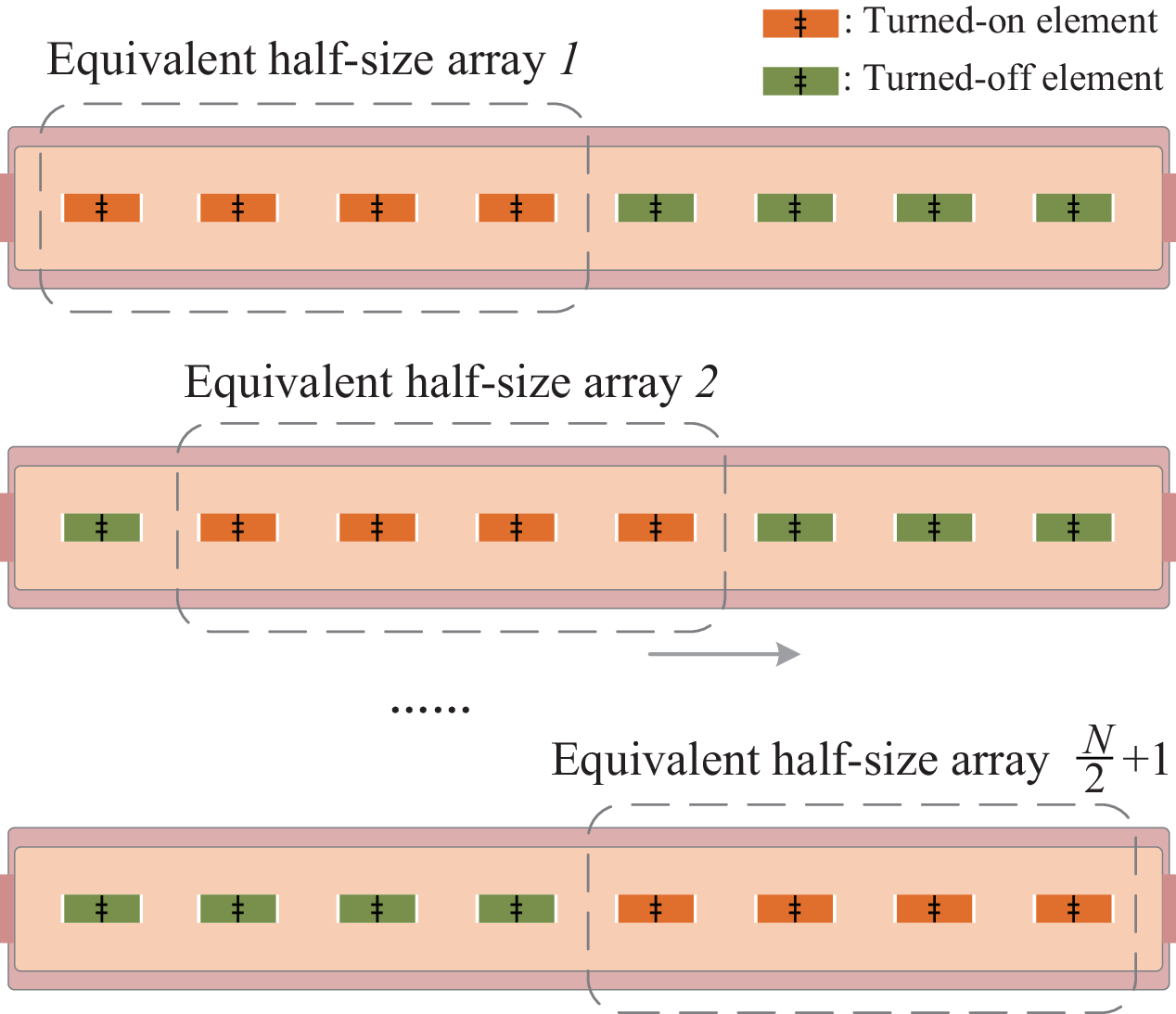}
	\caption{The RHS serves as a scale-changeable array.}
	\label{Sec2-Fig:Equivalent_array}
	\vspace{-0.5cm}
\end{figure}

Due to the amplitude-controllable property of the RHS, the element with an amplitude of $0$ essentially simulates the effect of being turned off. Therefore, \emph{it is feasible to configure an equivalent RHS array by activating certain elements without additional switches and control mechanisms, which leads to a scale-changeable array.} For example, as shown in Fig.~\ref{Sec2-Fig:Equivalent_array}, when a half of elements are turned off, the dimension of the RHS is equivalent to being halved, i.e., the beamwidth can be doubled in the angular domain. Moreover, turning off a specific number of elements introduces degrees of freedom in terms of the array topology. For an RHS with $N$ elements, there exist $N/2 + 1$ equivalent arrays at various positions as shown in Fig.~\ref{Sec2-Fig:Equivalent_array}, providing the selection gain as well.

\subsubsection{Hierarchical Codebook Structure}
Before introducing the hierarchical codebook, we first introduce the signal model for the near-field region. Under the spherical wave assumption, the channel between the user and the $n$-th RHS element can be written as
\begin{equation}
	h_n=\beta e^{-j\frac{2\pi}{\lambda}r_n},
\end{equation}
where $\beta$ is the channel gain and $r_n$ is the distance between the user to the $n$-th element. Assume that the coordinate of the $n$-th RHS element is $(\delta_n d,0)$ with $d$ being the element spacing, and the user position is $(r\cos \theta,r\sin \theta)$. Here, $r$ is the distance from the user to the origin and $\theta$ is the azimuth angle.
Therefore, $r_n$ can be given as
\begin{equation}\label{sec2:channel}
	\begin{split}
		r_n&=\sqrt{r^2-2\delta_n d r\cos \theta + \delta_n^2 d^2}\\
		&\stackrel{(a)}{\approx} r- \delta_n d \cos \theta +\frac{\delta_n^2 d^2(1-\cos^2\theta)}{2r} \\
		&\stackrel{(b)}{=} r- \delta_n d \psi + \frac{\delta_n^2 d^2 \mu}{2},
	\end{split}
\end{equation}
where the approximation (a) is obtained through the Taylor series expansion, while the transformation~(b) is obtained by letting $\psi=\cos\theta$ and $\mu=\frac{1-\cos^2\theta}{r} $.

With the channel characterization by parameters $\psi$ and $\mu$ in~\eqref{sec2:channel}, we design the codebook in the $\psi-\mu$ domain below. To serve users in different angles, the coverage of the BS spans $[-1,1]$ in $\psi$-domain.
Assume that the minimum distance between the user and the BS is $r_{min}$, the coverage of the BS in $\mu$-domain is  $[0, \frac{1}{r_{min}}]$ by calculating $\mu$. Thus, the hierarchical codebook covers both near-field and far-field regions.

\begin{figure}[t]
	\centering
	\includegraphics[width=0.45\textwidth]{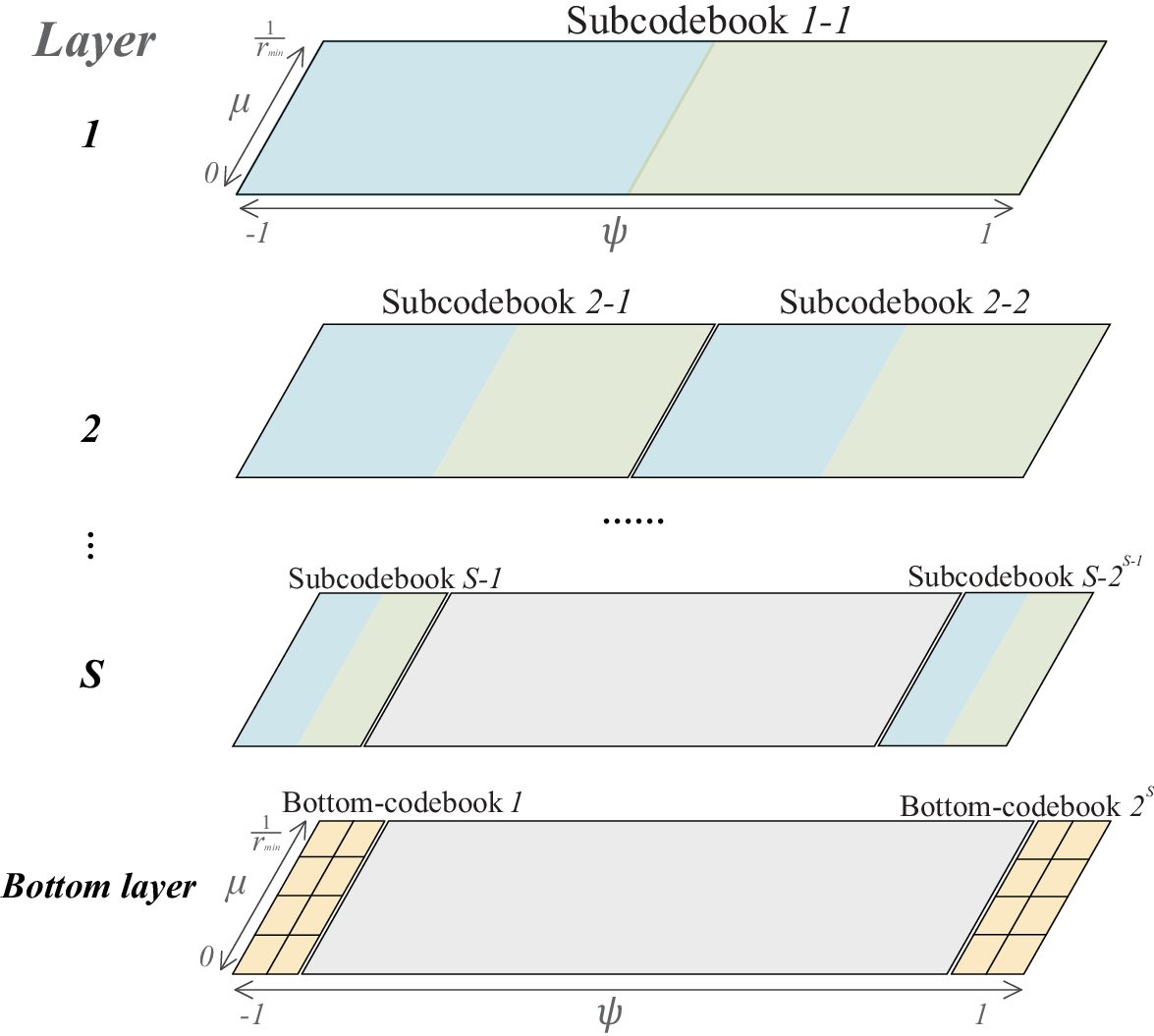}
	\caption{Hierarchical codebook structure for the RHS.}
	\label{Sec2-Fig:codebook_structure}
\end{figure}

As shown in Fig~\ref{Sec2-Fig:codebook_structure}, the hierarchical codebook is composed of the $S$ upper layers and a bottom layer. By extending the aperture of the equivalent array, the resolution of the upper-layer subcodebooks increases in $\psi$-domain, indicating the user's direction within the angular domain. Subsequently, the bottom-layer subcodebooks are designed to determine the user's position in $\mu$-domain. The design details can be found in \cite{SBHH-2024}.

Based on the subcodebooks designed above, the beam training can be performed in $\psi-\mu$ domains. To be specific, the upper-layer subcodebooks are first applied to reduce the search region in the $\psi$-domain. Then, the corresponding bottom-layer subcodebook is selected to determine the final positions in $\psi$-$\mu$ domains.

\subsubsection{Simulation Results}
We evaluate the performance of the proposed hierarchical codebook for the RHS. The number of the RHS elements is $N= 128 \times 8$ with $1$-bit amplitude control. For comparisons, we also obtain the performance results with the following different array structures and codeword design methods.
\begin{itemize}
\item \emph{Hierarchical single-beam phased-array codebook:} A single-beam codebook with the same size is implemented by a $64 \times 4$ phased-array, where each antenna of the phased-array is equipped with an additional switch in the simulation.
\item \emph{Hierarchical multi-beam phased-array codebook:} A multi-beam codebook with the same size implemented by a $64 \times 4$ phased-array.
\item \emph{Near-field phased-array codebook:} The codebook consists of 256 single-beam codewords implemented by a $64 \times 4$ phased-array.
\item \emph{Near-field RHS codebook:} The codebook consists of 256 single-beam codewords implemented by a scale-fixed RHS.
\end{itemize}

\begin{figure}[t]
	\centering
	\subfigure[]
	{
		\label{Sec2-Fig:sumrate}
		\includegraphics[width=0.4\textwidth]{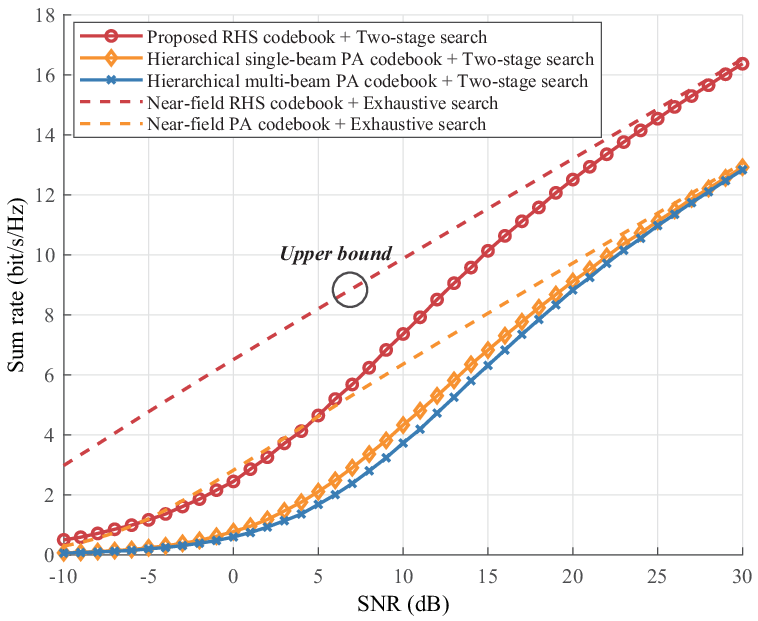}
	}
	\subfigure[]
	{
		\label{Sec2-Fig:overhead}
		\includegraphics[width=0.4\textwidth]{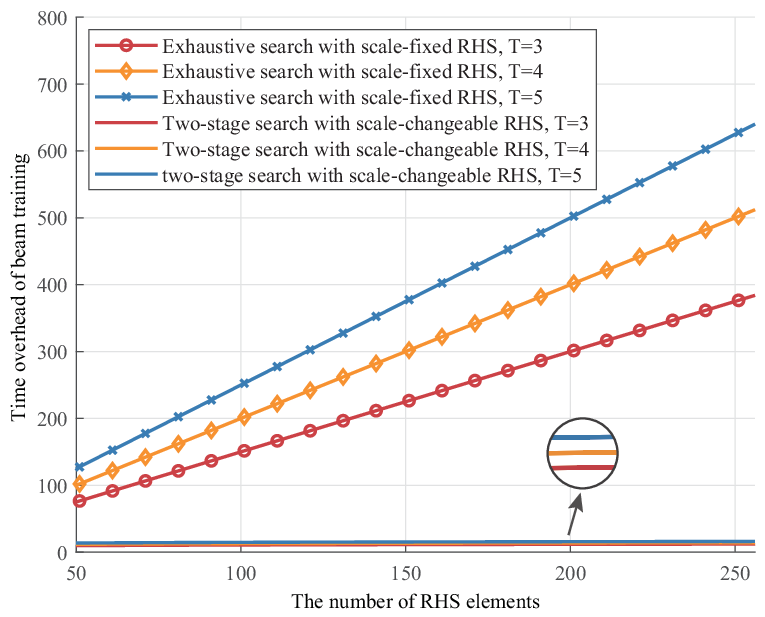}
	}
	\caption{Performance Analysis: (a) Sum rate vs. SNR; and (b) Beam training overhead vs. the number of the RHS elements.}
	\vspace{-0.5cm}
\end{figure}

Fig.~\ref{Sec2-Fig:sumrate} presents the sum rate performance with different values of the SNR. The sum rate of the  near-field RHS codebook with a exhaustive search scheme can be seen as the upper bound. It can be observed that the sum rate of the proposed scheme approaches the upper bound at the high SNR region. Moreover, we can observe that the single-beam phased-array codebook outperforms multi-beam phased-array codebooks due to the high and flat gain within the single-beam coverage. Fig.~\ref{Sec2-Fig:overhead} shows the time overhead of the beam training with different values of the number of RHS elements and the sampled step $T$ in the $\mu$ direction. With the proposed scale-changeable RHS, the overhead of the  beam training is reduced to a logarithmic level of the antenna number, while that of a scale-fixed RHS suffers a linear overhead of the number of RHS elements. In summary, compared to a scale-fixed RHS, the scale-changeable RHS can lead to a significant overhead reduction with a close sum rate performance.

\section{RHS-aided Wireless Sensing}
\label{Sec:4}
Future wireless networks are envisioned to provide flexible and seamless sensing capability~\cite{HBKZHL-2022}, in order to support emerging applications, such as indoor navigation \cite{EMOSAGMKYM-2018}, healthcare \cite{HDYJYS-2017}, and theft detection \cite{JQMY-2018}. The basic idea of utilizing wireless signals for sensing purposes is to extract information from the changes of wireless signals. To be specific, the presence or movement of objects will result in a variation of certain properties (e.g., phase or amplitude) of received signals~\cite{JHYYC-2020}. Through analyzing the received signals, the receiver can recognize the movement of users or the existence of objects. 

A commonly-used scheme for wireless sensing applications is the millimeter-wave (mmWave) radar \cite{KMVBS-2019}, which utilizes a phased-array to generate directional beams to detect the targets. As we have introduced before, the phased-array has a complicated circuit and its energy costs are high, which limits the performance. The RHS has shown a potential to address the above issues, and thus the RHS radar is a promising alternative to the radar systems enabled by the phased-array. 

In this section, we first introduce the key techniques for the RHS radar, and then present how to integrate such a RHS radar system with communication systems, leading to a holographic ISAC system.

\subsection{RHS Radar}
\label{sec:rhs_radar}

\begin{figure}[!t]
	\centering
	\includegraphics[width=0.45\textwidth]{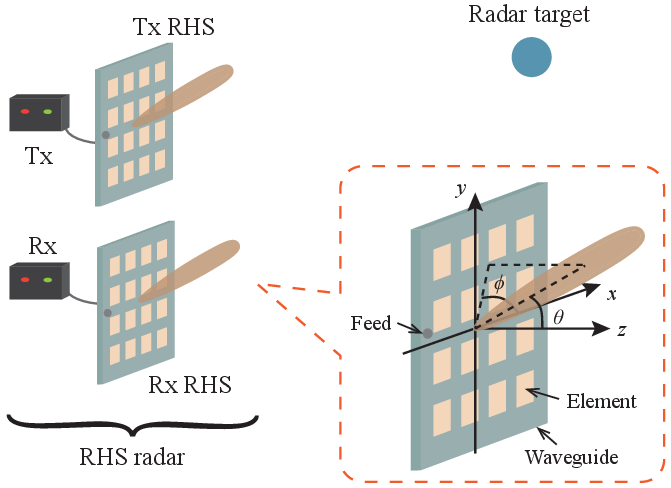}
	\caption{An RHS radar system with a single target.}
	\label{Sec3-Fig:f_scenario}
\end{figure}

An RHS radar system is composed of a single radar target, a Tx with an RHS, and an RX with an RHS, as shown in Fig.~\ref{Sec3-Fig:f_scenario}. The target is assumed to be located in the far-field of the radar system. Let $(R,\theta,\phi)$ be the position of the target, where $R$ is the distance between the Tx and the target, $\theta$ and $\phi$ are the azimuth and evaluation angles, respectively. We also denote $\beta \in \mathbb{C}$ as the reflection coefficient of the target. The tasks of a radar system can be categorized into two types:
\begin{itemize}
	\item \emph{Target Detection:} In this type of task, the radar is to detect the existence of the target at the direction of $(\theta,\phi)$;
	
	\item \emph{Parameter Estimation:} In this type of task, the radar is to estimate a certain parameter, for example, the reflection coefficient of the target $\beta$ and the distance $R$.
\end{itemize}
In the following, we will show the models for both types of tasks.

\subsubsection{System Model}
Let $\bm{M}^t$ and $\bm{M}^r$ be the beamformer matrices of the Tx RHS and the Rx RHS, respectively. Here, $\bm{M}^t$ and $\bm{M}^r$ are diagonal matrices whose $n$-th diagonal element is the amplitude response of the $n$-th element of the corresponding RHS. Let $\bm{a}^r(\theta,\phi)$ and $\bm{a}^t(\theta,\phi)$ be the steering vectors for the Tx RHS and the Rx RHS, respectively. We also define $\bm{h}^t$ and $\bm{h}^r$ as the channel from the Tx to the target and the channel from the target to the Rx, respectively. Let $x(t)$ be the transmitted signals, $\bm{q}^t$ and $\bm{q}^r$ are the phase vectors of the Tx RHS and the Rx RHS, respectively. With these notations, the received signal can be expressed by \cite{HHBZL-2023} 
\begin{equation}
	\label{echo-signal}
	\begin{aligned}
	y(t) &= \beta (\bm{M}^r \bm{q}^r)^T (\bm{a}^r(\theta,\phi))^T  (\bm{h}^r)^T \bm{h}^t \bm{a}^t(\theta,\phi)\bm{M}^t \bm{q}^t x(t) \\
	&~~~+ (\bm{M}^r \bm{q}^r)^T\bm{n},\\
	&= r(t) + \tilde{n}(t),
	\end{aligned}
\end{equation}
where $\bm{n}$ is the noise. Here, $r(t)$ is the echo signals from the target and $\tilde{n}(t)$ is the equivalent noise.

\textbf{Target Detection:} For the target detection tasks, we consider a binary hypothesis test model to detect whether there exists a target in a certain direction \cite{XHHB-2023}. To be specific, we have two hypotheses: 1) hypothesis $\mathcal{H}_0$ indicates that there exists no targets; and 2) hypothesis $\mathcal{H}_1$ indicates that there exists a target. The receiver will detect the energy of received signals for a certain period and compare it with a detection threshold. The receiver will take hypothesis $\mathcal{H}_0$ if the energy is lower than the threshold, and otherwise take hypothesis $\mathcal{H}_1$. 

Following the Neyman-Pearson criterion, the optimization objective is to maximize the detection probability given the false alarm probability. As the detection probability is positively related to the received signal-to-noise ratio (SNR) given the given the false alarm probability. Therefore, the objective can be reformulated as the maximization of the SNR.

\textbf{Parameter Estimation:} For the parameter estimation tasks, we use the maximum likelihood estimator to estimate the parameters \cite{XHHLB-2022}, whose performance is typically measured by the Cram\'er-Rao bound (CRB). The CRBs of the unknown parameters, for example, the reflection coefficients and the distance, are all inversely proportional to the SNR. Therefore, the performance can also be optimized by maximizing the SNR.

\subsubsection{Amplitude Control Method}
For the above two tasks, the optimization problem can be written as
\begin{equation}
	\begin{aligned}
		\max \limits_{\bm{M}^t,\bm{M}^r}~&\gamma\\
		s.t.~& \|\bm{M}^t\bm{q}^tx^t\|^2 = P^t,
	\end{aligned}
\end{equation}
where $\gamma$ is the received SNR as expressed below 
\begin{equation}
	\gamma = \beta \underbrace{\dfrac{|(\bm{M}^r \bm{q}^r)^{T} (\bm{a}^r (\theta, \phi))^{T}(\bm{h}^r)^T|^2 }{|\bm{M}^r \bm{q}^r|^2 \sigma^2}}_{\gamma^r} \cdot \underbrace{|\bm{h}^t\bm{a}^{T}_t (\theta, \phi) \bm{M}^t \bm{q}^t x|^2}_{\gamma^t}.\label{def_gamma}
\end{equation} 
The constraint is the power conservation brought by the RHS.

As the optimization variables $\bm{M}^t$ and $\bm{M}^r$ for Tx and Rx RHSs can be optimized separated since they are decoupled in the constraints. Therefore, transmit and receive subproblems can be written as the following form
\begin{equation}
	\begin{aligned}
		\max_{\bm{M}}~&|\bm{a}^{T} \bm{M} \bm{q}|^2, \label{p4_obj}\\
		s.t.~&|\bm{M} \bm{q}|^2 = 1. 
	\end{aligned}
\end{equation}
A closed form optimal solution can be found for such a problem \cite{HHBZL-2023}. It should be noted that the above results are obtained under the assumption that the amplitude for each element is continuous. In practical systems, the available amplitudes for each element are discrete. We need to quantize the optimized continuous results to obtain the discrete ones.

\subsubsection{Simulation Results}
\begin{figure}[t]
	\centering
	\includegraphics[width=0.45\textwidth]{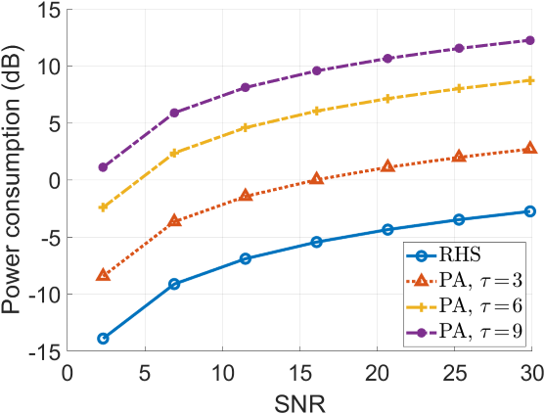}
	\caption{Performance analysis: Power consumption vs. Received SNR}
	\label{Sec3-Fig:Simulation}
\end{figure}

\begin{figure*}[!t]
	\centering
	\subfigure[]{
		\includegraphics[width=0.45\textwidth]{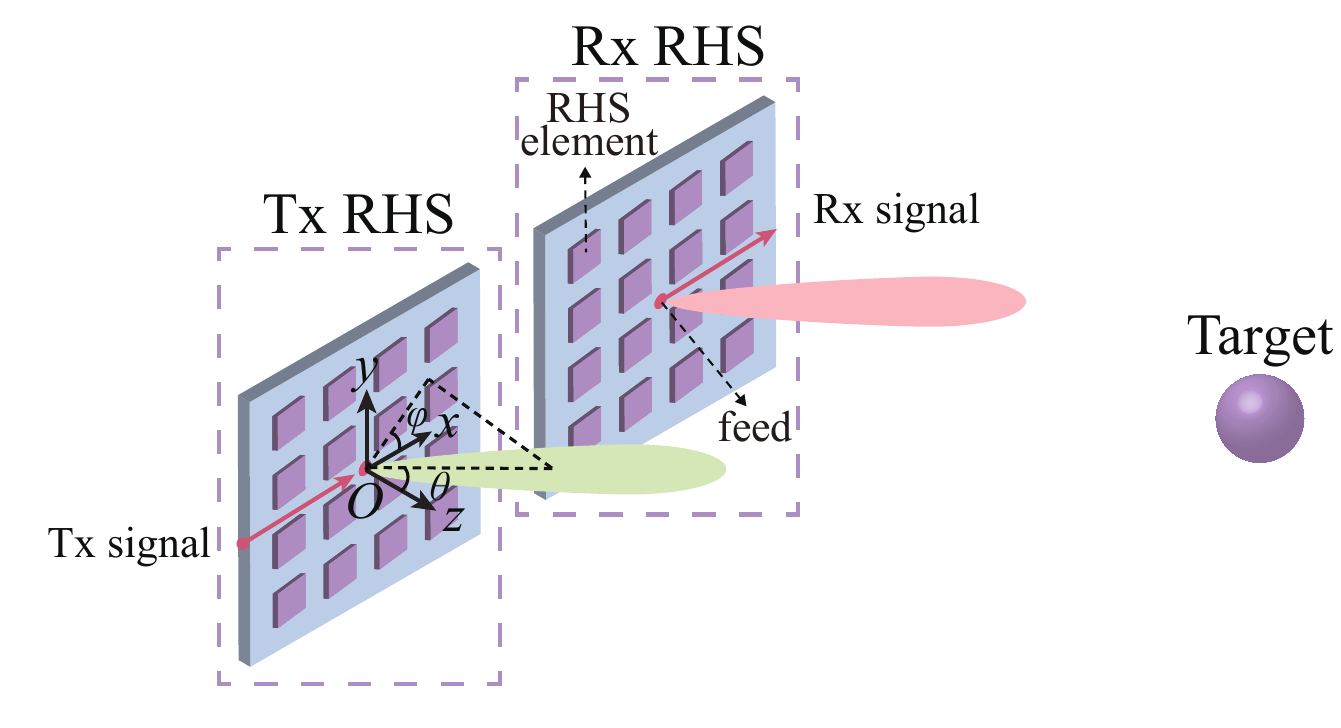}
	}
	\subfigure[]{
		\includegraphics[width=0.45\textwidth]{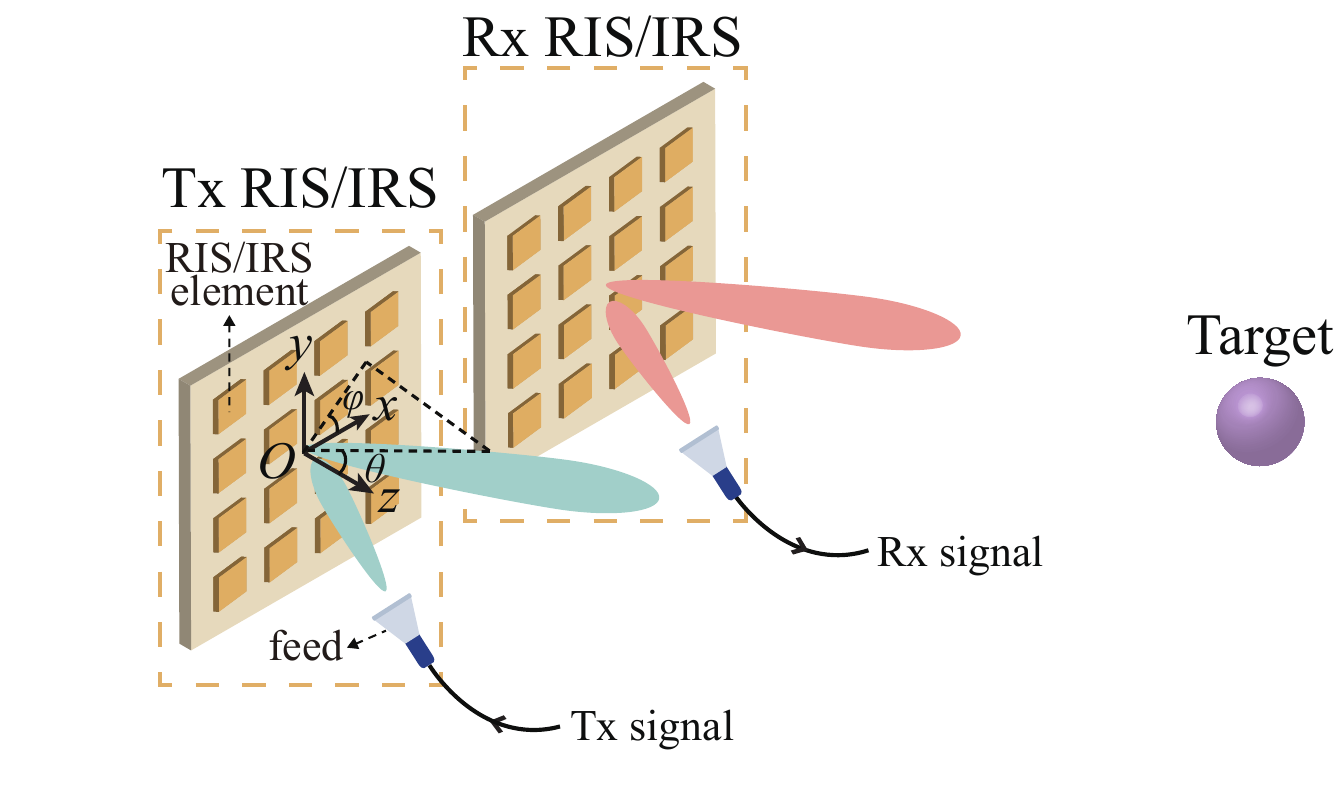}
	}
	\caption{System Model: (a) RHS radar systems; (b) RIS/IRS radar systems (3D view).} 
	\label{fig_dis}
\end{figure*}

In Fig. \ref{Sec3-Fig:Simulation}, we show the total power consumption versus the received SNR of the RHS radar and the phased-array radar given the same hardware cost. $\tau$ is the cost ratio of a phased-array antenna to an RHS element, which is set as $\tau = 3, 6, 9$, respectively \cite{P-2019}. We can observe that the power consumption of both RHS radar and phased-array radars increases as the SNR increases. Moreover, the power consumption of the RHS radar is lower than that of the phased-array radar, which shows the energy efficiency of the RHS radar.

\subsection{Comparison: RHS Radar or RIS/IRS Radar?}
After introducing the RHS radars, we will make a comparison with the RIS/IRS radars \cite{HBKZHL-2022} to show the condition when the RHS radars outperform the RIS/IRS radars. In this part, we take a single target detection as a simple example to obtain the physical insights. The radar system consists of a Tx, a Rx, and two co-located surface antennas, each serving as a transmit antenna or a receive antenna. A target is located at unknown range $R$ and direction $(\theta,\varphi)$ in the far-field of the radar, where $\theta$ and $\varphi$ denote the elevation and azimuth angles, respectively. The aim of the radar system is to determine the direction angles $(\theta,\varphi)$ of the target and estimate the target range $R$ \cite{XHLZHB-2025}. The angles and range of the target are determined by analyzing the received signals from each direction of interest and its corresponding time delay. As shown in Fig. \ref{fig_dis}, we use an RHS and an RIS/IRS to construct two radar systems. In the RHS radar system, the feed that connects Tx or Rx is embedded in the RHS, while in the RIS/IRS radar system, the feed is positioned in the near-field of the surface.

\subsubsection{Signal Model and Beamformer Design}
For the RHS radar, the signal model and beamformer design have been introduced in Section \ref{sec:rhs_radar}. Similar to the RHS radar introduced in (\ref{echo-signal}), the echo signal for the RIS/IRS radar can be expressed as
\begin{equation}
		y(t) = \beta (\bm{h}^r)^T (\bm{\Psi}^r)^T \bm{H} \bm{\Psi}^t\bm{h}^t x(t) + n,\\
\end{equation}
where $\bm{h}^r$ and $\bm{h}^t$ are the feed - RIS/IRS channels for Tx and Rx, respectively, $\bm{H}$ is the channel gain between the Tx RIS/IRS and the Rx RIS/IRS containing the reflection of the target, $\bm{\Psi}^t$ and $\bm{\Psi}^r$ are the beamformer matrices for Tx RIS/IRS and Rx RIS/IRS, respectively. Different from the RHS, each element of the RIS/IRS can be modeled as a phase shifter. Mathematically speaking, the diagonal element of $\bm{\Psi}^t$ and $\bm{\Psi}^r$ can be expressed as 
\begin{equation}
	\psi_n = A e^{j\theta_n},
\end{equation}
where $A$ is the constant reflection amplitude and $\theta_n$ is the phase shift of the reflected signal compared to the incident signal. Therefore, the SNR of the echo signals can be expressed as
\begin{equation}\label{SNR_RIS_prec}
	\gamma^{RIS} = \frac{|\beta (\bm{h}^r)^T (\bm{\Psi}^r)^T\bm{H}\bm{\Psi}^t\bm{h}^{t}|^2 P_t}{\sigma^2},
\end{equation}
where $P_t = |x(t)|^2$ is the transmit power of signals at the feed. With the aim to maximize the SNR, the optimal phase of each RIS/IRS element should be equal to the negative value of the phase of the channel gain corresponding to the RIS/IRS element so that the phases of all paths can be aligned at the Rx. More details can refer to \cite{XHLZHB-2025}.

\subsubsection{Performance Analysis}
In this part, we will start from the 1D case and then show the results for the 2D case.

\begin{figure}[!t]
	\centering
	\includegraphics[width=0.45\textwidth]{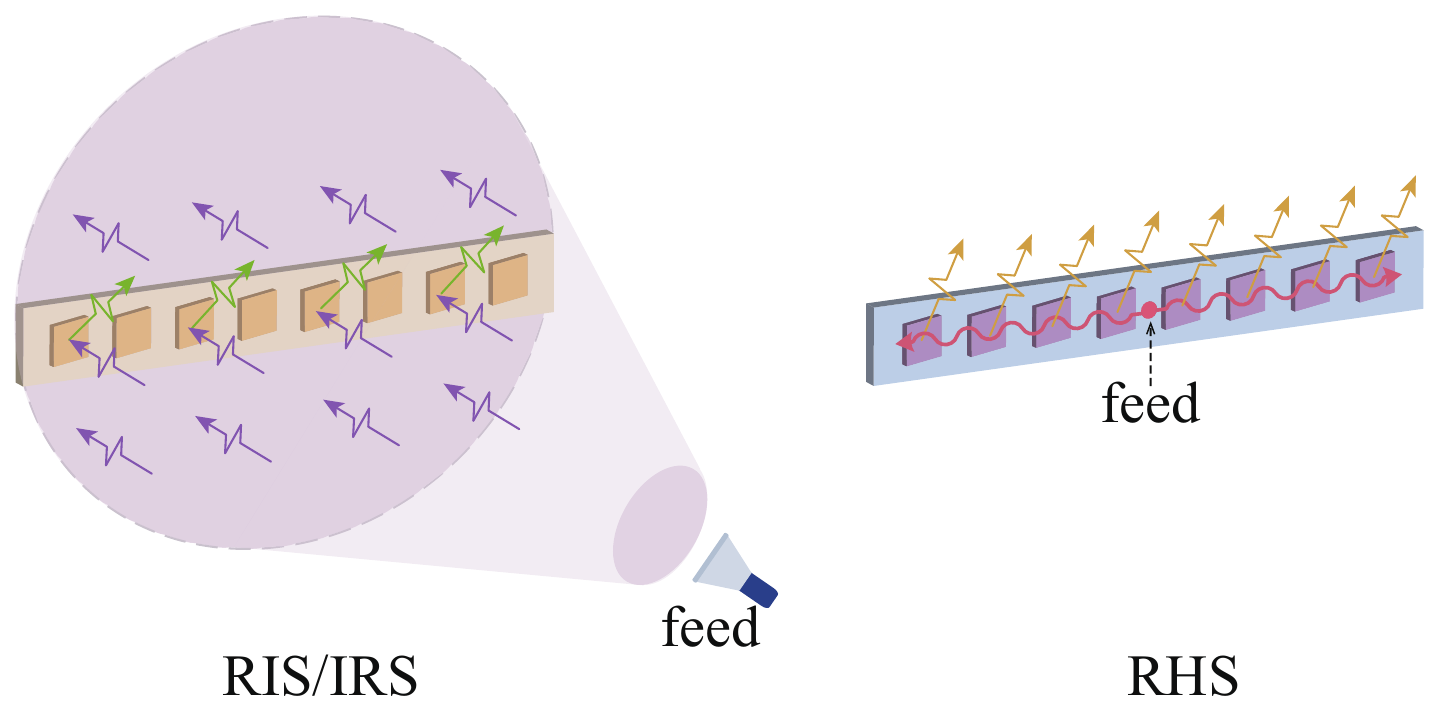}
	\caption{Schematic of the RIS/IRS and the RHS in the 1D case.} 
	\label{fig_1D_demons}
\end{figure}

\textbf{1D Case:} Our results in \cite{XHLZHB-2025} reveals that when the operating frequency is higher than a threshold $f_{th}$, the SNR upper bound of the RHS radar is always higher than that of the RIS/IRS radar when their physical lengths approach to infinity. This observation can be explained as shown in Fig. \ref{fig_1D_demons}. Note that the physical dimension of the surface is related to the wavelength. For the RIS/IRS radar, only a small portion of the signal energy emitted from the feed can be reflected if the operating frequency is larger than a threshold, since the physical width of the RIS/IRS is too small to capture most of the energy. However, for the RHS radar, the signals propagates along the whole surfaces, thus the signal energy can be primarily transferred to the RHS surface from the feed. This means that the energy loss of the RHS radar is much smaller than that of the RIS/IRS radar. 

\begin{table}[!t]
	\centering \caption{The Optimal Type of the Surface Under Various Conditions}
	\begin{threeparttable} 
		\begin{tabular}{| c | c | c | c | c |}
			\hline \diagbox{frequency}{size}   & $16\times 16$     & $32\times 32$     & $64\times 64$ & $128\times 128$  \\
			\hline\hline  $1$ GHz &  \cellcolor{gray!30} I &  \cellcolor{gray!30} I &  \cellcolor{gray!30} I &  \cellcolor{gray!30} I \\
			\hline\hline  $2.4$ GHz  & \cellcolor{gray!30} I & \cellcolor{gray!30} I & \cellcolor{gray!30} I & \cellcolor{gray!30} I   \\
			\hline\hline  $12$ GHz  & H & H & \cellcolor{gray!30} I &\cellcolor{gray!30} I   \\
			\hline\hline  $24$ GHz  & H & H & H & H   \\
			\hline\hline  $77$ GHz  & H & H & H & H   \\
			\hline
		\end{tabular}
		\begin{tablenotes} 
			\item Note: H represents RHS; I represents RIS/IRS. 
		\end{tablenotes} 
	\end{threeparttable} 
	\label{table_comp}
\end{table}

\textbf{2D Case:} For the 2D case, the conclusion will be different. With a increased aperture, the SNR of the RIS/IRS radar increases since more energy can be reflected. But for the RHS radar, as we have mentioned before, there exists the energy loss when signals propagate along the surface, the energy loss of the 2D RHS will also increase. Therefore, it is possible that the SNR of the RIS/IRS radar surpasses that of the RHS radar when the physical sizes of the surfaces are large enough. In Table \ref{table_comp}, we show the optimal type of the surface (RIS/IRS or RHS) under different working frequencies and sizes. We can observe that there exists a size threshold under which the RHS radar outperforms the RIS/IRS radar, and the size threshold is an increasing function of the working frequency.

\subsubsection{Simulation Results}
\begin{figure}[!t]
	\centering
	\subfigure[]{
		\includegraphics[width=0.40\textwidth]{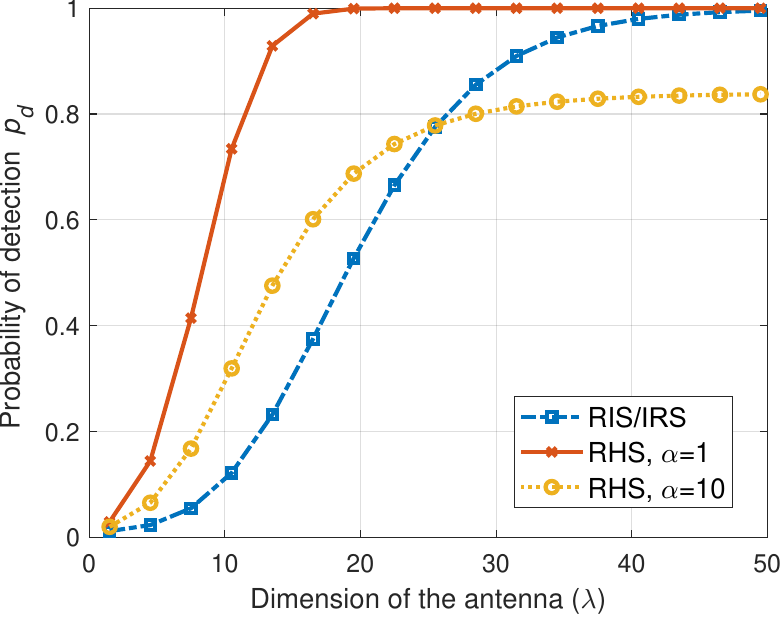}
	}\\
	\subfigure[]{
		\includegraphics[width=0.40\textwidth]{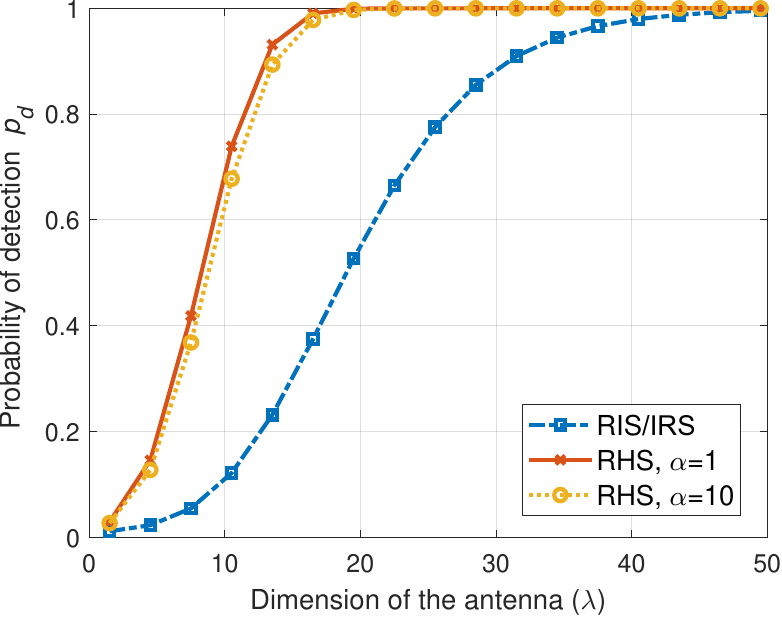}
	}
	\caption{Probability of detection $p_d$ vs. the physical width of the antenna when (a) $f=10$ GHz; (b) $f=30$ GHz.} 
	\label{fig_compare_2D}
\end{figure}
Fig.~\ref{fig_compare_2D} shows the probability of detection versus the physical width of 2D surfaces with a fixed physical length (along the $x$-axis). The frequency of the radar signal is set as $10$ GHz in Fig.~\ref{fig_compare_2D}(a) while it was set as $30$ GHz in Fig.~\ref{fig_compare_2D}(b). It can be observed that when the radar was operated in $30$ GHz, the probability of detection $p_d$ of the RHS radar is higher than that of the RIS/IRS radar. When the working frequency is $10$ GHz, $p_d$ of the RIS/IRS radar are inferior than that of the RHS radar when the physical width is small, and is superior than that of the RHS radar when the physical width exceeds a threshold. The simulation results verify our performance analysis.

\subsection{Holographic Integrated Sensing and Communications}
We have presented two study cases for the RHS radars for sensing applications in previous parts. However, it is also important to integrate the sensing functions with current communication systems to save the hardware costs. In the following, we will show how the RHS works for an ISAC system \cite{HHBMZHL-2022}.

\begin{figure}[!t]
	\centering
	\includegraphics[width=0.45\textwidth]{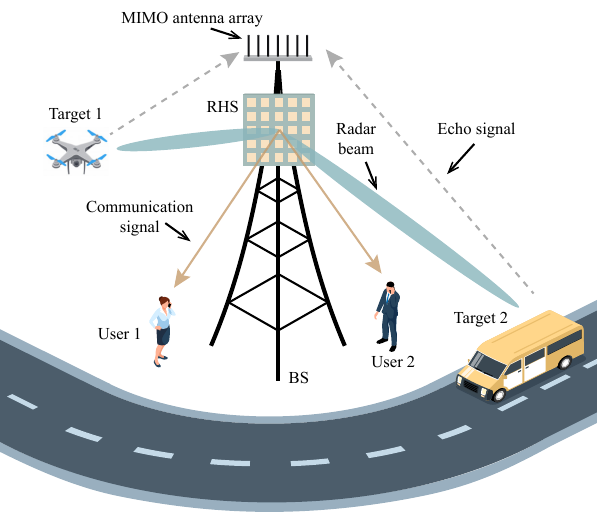}
	\caption{A holographic ISAC system.}
	\label{f_scenario}
\end{figure}

A holographic ISAC system is shown in Fig.~\ref{f_scenario}, which consists of $L$ mobile users, multiple targets, and a BS equipped with an RHS as a transmit antenna and a MIMO antenna array as a receive antenna\footnote{The RHS can serve as the receive antenna as well. In this case study, we will highlight how the RHS changes the transmit waveform design and use traditional MIMO antenna array in the receiver side.}. To perform the ISAC functions, the following three steps are executed sequentially.

\begin{itemize}
		\item \textbf{Optimization:} The BS optimizes the ISAC signals to maximize the radar sensing performance while guaranteeing the quality of service (QoS) for communication users.
		
		\item \textbf{Transmission:} The RHS transmits ISAC signals with multiple beams towards the directions of users and targets.
		
		\item \textbf{Reception:} The communication users receive the signals from the RHS and decode the received signals. At the same time, the MIMO antenna array listens to the echo signals reflected by the targets for radar sensing. 
\end{itemize}

\subsubsection{Holographic Beamforming Optimization}
To generate beams for both communication and sensing, we utilize the holographic beamforming diagram introduced in Section \ref{Sec:holographic-beamforming}, where the RHS is used for the analog beamforming. It should be noted that the radar waveforms and the communication data streams will be combined and fed into the digital beamformer to achieve ISAC functions. The following metrics are utilized for sensing and communication:
\begin{itemize}
	\item \textbf{Sensing:} According to the results in~\cite{PJY-2007}, the radar performance is related to the beampattern gains toward desired directions and the cross-correlation among these directions. Specifically, by increasing the beampattern gains in correspondence of the target directions, the signal-to-interference-and-noise ratio (SINR) of echo signals can be improved, leading to a higher sensing performance. Moreover, for the multi-target detection, it becomes easier to separate the echo signals from different directions by decreasing the cross-correlation, which enhances the sensing accuracy as well. Therefore, the radar utility function can be defined as a linear function of the beampattern gains and the cross-correlation among different directions.
	
	\item \textbf{Communication:} We simply use the SINR of each user to quantify the performance of communication.
\end{itemize} 
Therefore, the ISAC optimization problem is to maximize the radar utility given SINR and power constraints by optimizing digital and analog beamformers. This problem can be solved by alternating optimization of digital and analog beamformers. More details can be found in \cite{HHBMZHL-2022}. 

\subsubsection{Cost-effectiveness Analysis}
The cost-effectiveness is defined as the cost saving of using an RHS instead of using a traditional phased array. For fairness, we assume that both systems have the same radar utility function $\delta$. We utilize the following effectiveness metric~\cite{NNNXJF-2016}:
\begin{equation}
	\eta(\delta) = 1 - \alpha_r(\delta)/\alpha_a(\delta),\label{def_eta}
\end{equation}
where $\alpha_r(\delta)$ is the cost of the RHS to obtain the radar utility $\delta$, and $\alpha_a(\delta)$ is the cost of the MIMO array to obtain the same utility $\delta$. Specifically, the cost of the RHS is lower than the cost of the MIMO array if $\eta(\delta) > 0$, and a greater cost saving is achieved with a larger value of $\eta(\delta)$.

For the cost, assume that the hardware cost of each RHS element is $\nu$. Then, the cost of an RHS is $\alpha_r(\delta) = \nu M + \chi K$, where $\chi$ is the cost of one RF chain. Similarly, the hardware cost of a phased array is $\alpha_a(\delta) = \beta \nu M_A + \chi K$, where $\beta$ denotes the ratio of the antenna cost of a phased array to the cost of an RHS element.

For a phased array, we have $\delta = M_A P_M$ \cite{R-2017}, and its cost is given as $\alpha_a(\delta) = \dfrac{\beta \nu \delta}{P_M} + \chi K$. As for the RHS, according to the results in \cite{HHBMZHL-2022}, the cost-effectiveness can be approximated as follows:
\begin{equation}
	\eta(\delta) \ge 1 - \dfrac{6 \nu \delta + P_M \chi K}{\beta \nu \delta + P_M \chi K}.
\end{equation}

If $\beta > 6$, we can infer that $\eta(\delta)$ is always greater than $0$ for any $\delta$, which provides the condition under which an RHS is more cost-effective than a phased array. As phased arrays usually require many expensive components, such as phase shifters and power amplifiers for beam steering, the cost ratio is much greater than $6$~\cite{P-2021}.

\subsubsection{Simulation Results}

 \begin{figure}[t]
	\centering
	\includegraphics[width=0.45\textwidth]{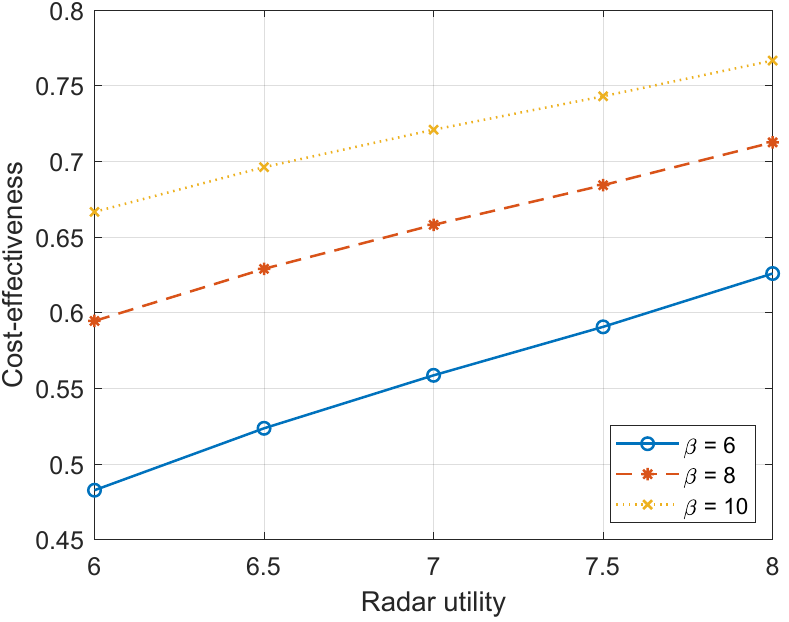}
	\caption{Cost-effectiveness $\eta$ versus the radar utility $\delta$ for different values of cost ratio $\beta$.}
	\label{f_cost}
\end{figure}

Fig.~\ref{f_cost} illustrates the cost-effectiveness $\eta$ of the proposed scheme with respect to the phased array as a function of the radar utility $\delta$ for different values of cost ratio $\beta$ when $\rho = 0.8$. A positive value of cost-effectiveness implies that the cost of the RHS is smaller than that of the phased array, and a larger value of the cost-effectiveness implies that a larger cost saving in favor of the RHS. We observe that, for any values of $\beta$, the cost-effectiveness is always greater than $0$. Besides, $\eta$ increases with the radar utility $\delta$, which indicates that larger cost savings can be obtained with the proposed scheme when a higher performance requirement of radar sensing is imposed.

\section{RHS Implementation and Prototypes}
\label{Sec:5}
In this section, we present the hardware implementation of the RHS and its prototypes. In Section \ref{sec:implementation}, we show the hardware design of the RHS. After the implementation of the RHS, we build the prototype of RHS aided wireless communications and ISAC systems in Sections \ref{sec:wireless_prototype} and \ref{sec:ISAC}, respectively.

\subsection{RHS Implementation}
\label{sec:implementation}
The implementation of the RHS follows the logic from the element to the array. In the following, we will first show the RHS element structure and then introduce how to assemble the elements into an antenna array.

\subsubsection{Element Design}
\begin{figure}[t]
	\centering
	\subfigure[]{
		\centering
		\includegraphics[width=0.4\textwidth]{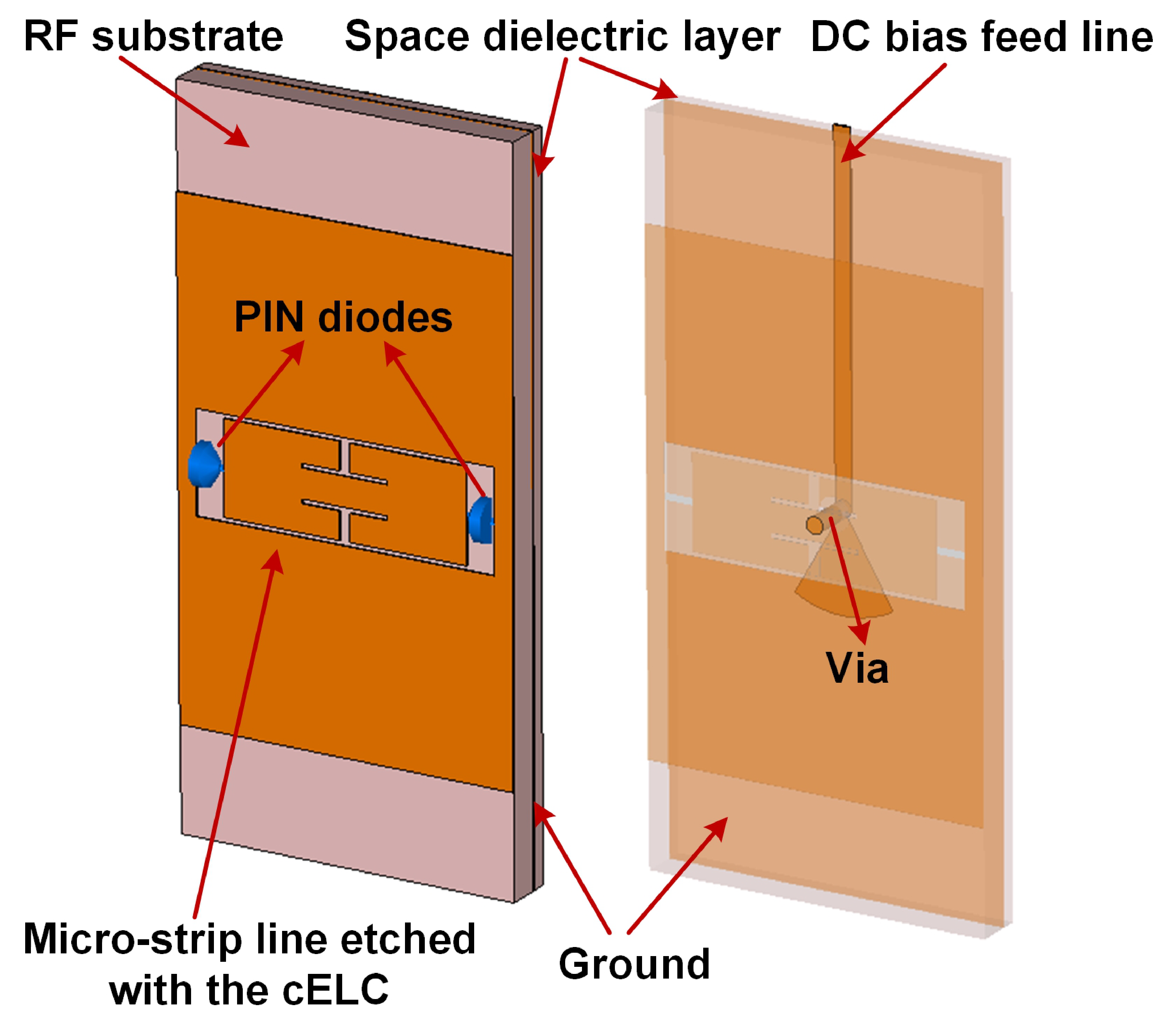}
	}
	\subfigure[]{
		\centering
		\includegraphics[width=0.4\textwidth]{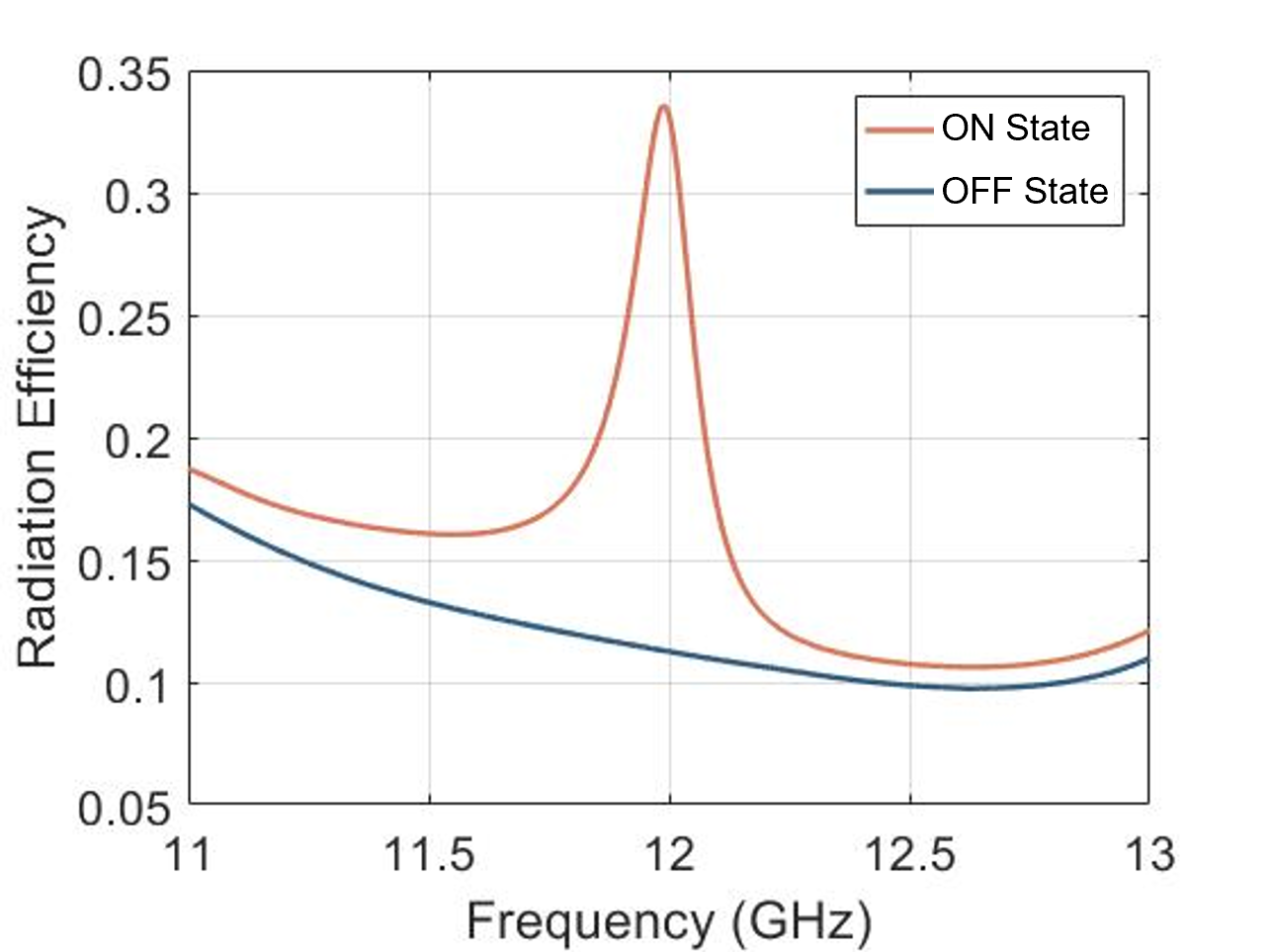}
	}
	\caption{Element design: (a) cELC-based structure; (b) Radiation power with different states of PIN diodes.}
	\label{comp}
\end{figure}

An RHS element capable of controlling the radiation amplitude of the reference wave generated by the feed relies on a complementary electric-LC (cELC) resonator loaded with PIN diodes~\cite{TMWJTMD-2016}. As shown in Fig. \ref{comp}(a), the whole structure of the RHS element consists of the following five layers:
\begin{itemize}
	\item \textbf{Top layer:} the micro-strip line etched with the cELC resonator.
	\item \textbf{2nd layer:} the radio frequency (RF) substrate to guide the propagation of reference waves.
	\item \textbf{3rd layer:} the ground plane made by copper.
	\item \textbf{4th layer:} the spacer dielectric layer to separate the ground plane and the direct-current (DC) bias feed line.
	\item \textbf{Bottom layer:} the DC bias feed line to apply biased voltage to the PIN diodes.
\end{itemize}
To improve its design freedom, a T-shaped slot is put in the middle of each long side of the rectangular metal patch to provide more adjustable geometric properties. The mutual inductance of the cELC resonator as well as the radiation power of the reference wave is determined by the states of embedded PIN diodes.

Fig. \ref{comp}(b) shows the full-wave analysis of the RHS element simulated in the CST Microwave Studio. In the simulation, the reference wave generated by port 1 radiates energy through the annular slot on the RHS element, and the residual energy of the reference wave is absorbed by port 2. It can be seen that the ratio of the radiated power when the PIN diodes are in the OFF states to that when the PIN diodes are in the ON states is about~3, indicating that the RHS element with a cELC resonator is capable of controlling the radiation amplitude of the reference wave~\cite{RYHBHHL-2023}.

It should be noted that the controllable radiation amplitude can be fabricated by tunable materials including PIN diodes, varactor diodes, and liquid crystal. The characteristics of these fabrication methods are summarized in Table \ref{RHS-compt}. We need to select suitable fabrication methodology of the RHS element for different applications. 
\begin{table*}[!t]
	\centering
	\renewcommand\arraystretch{1.6}
	\caption{Comparison Between Different Fabrication Methods}
	\begin{tabular}{|m{2.0cm}<{\centering}|m{2.0cm}<{\centering}|m{2.0cm}<{\centering}|m{2.0cm}<{\centering}|m{2cm}<{\centering}|m{2cm}<{\centering}|}
		\hline
		\textbf{Method}&\textbf{Tuning range} & \textbf{Precision} & \textbf{Integration} & \textbf{Control circuit} & \textbf{Energy loss} \\
		\hline
		\textbf{PIN diodes} & Wide& Binary control & Easy & Simple  & High\\
		\hline
		\textbf{Varactor Diodes}& Wide& Continuous control & Easy & Complicated  & High\\
		\hline
		\textbf{Liquid crystal}& Limited & Continuous control & Need to be packed & Complicated & Low \\
		\hline
	\end{tabular}
	\label{RHS-compt}
\end{table*}

\subsubsection{Array Design}
After the design of RHS elements, we also need to design the RHS array to maximize the radiation efficiency \cite{RYHBHL-2023}.

\begin{figure}[t]
	\centering
	\includegraphics[width=0.45\textwidth]{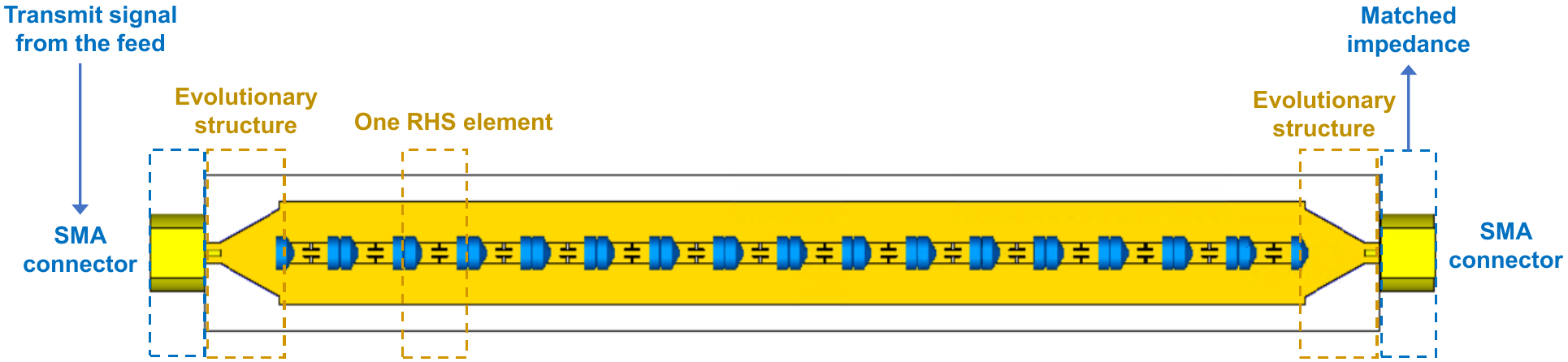}
	\caption{1D RHS Antenna Array.}
	\label{antenna}
\end{figure}

\textbf{1D RHS antenna array:} As shown in Fig. \ref{antenna}, a one-dimensional RHS antenna array consists of multiple elements, two Sub Miniature version A (SMA) connectors, and two symmetric evolutionary structures. To be specific, the SMA connector connects the feed which inputs the transmit signals, and the evolutionary structure is designed for impedance matching to maximize the energy imported into the antenna.

\begin{figure}[t]
	\centering
	\includegraphics[width=0.45\textwidth]{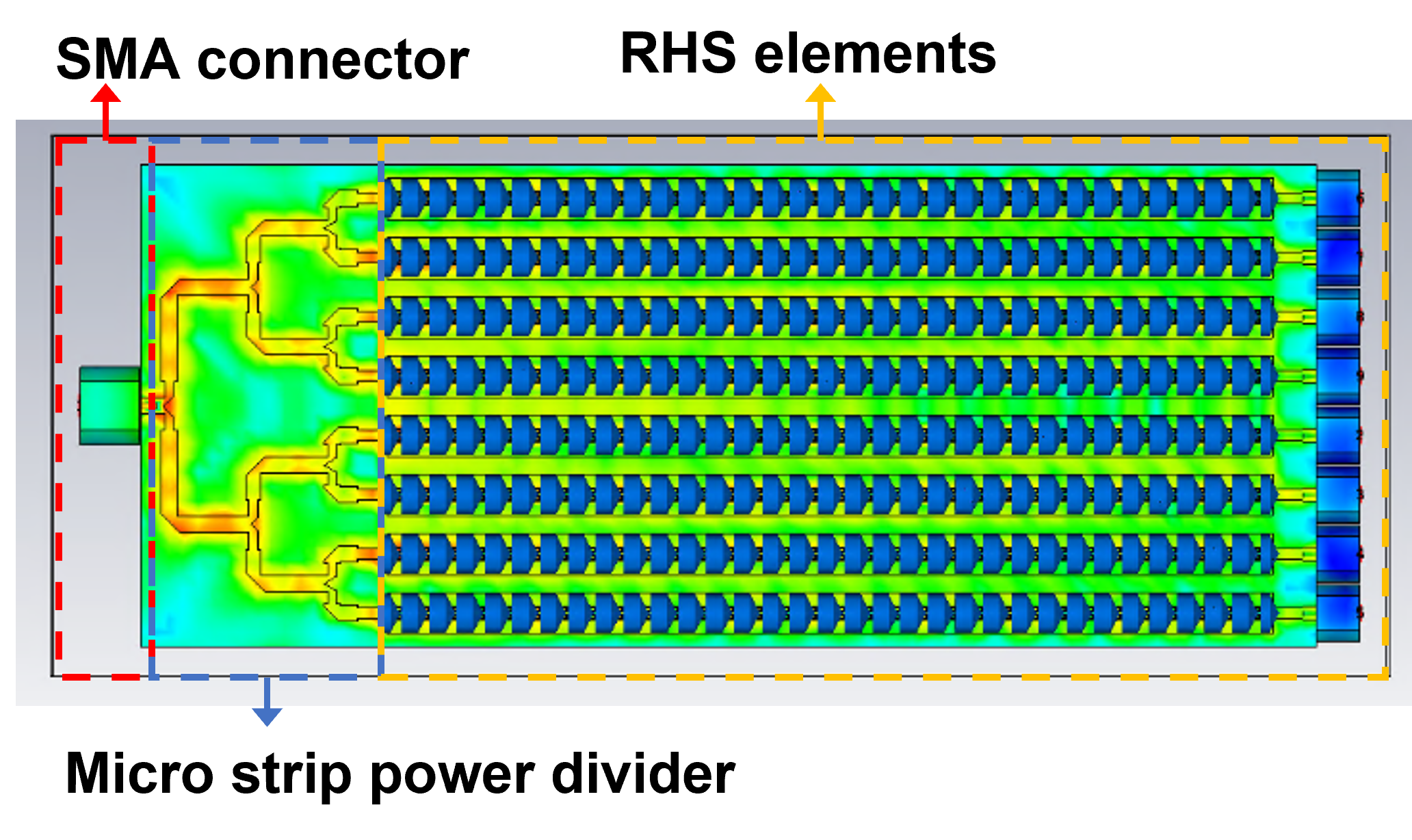}
	\caption{2D RHS Antenna Array.}
	\label{antenna2}
\end{figure}

\textbf{2D RHS antenna array:} As shown in Fig. \ref{antenna2}, a two-dimensional RHS antenna array consists of multiple elements, a SMA connector, and a micro strip power divider. It should be noted that different from the 1D case, the micro strip power divider is to deliver the reference waves input from the feed to the whole surface.  

\begin{figure*}[t]
	\centering
	\includegraphics[width=0.9\textwidth]{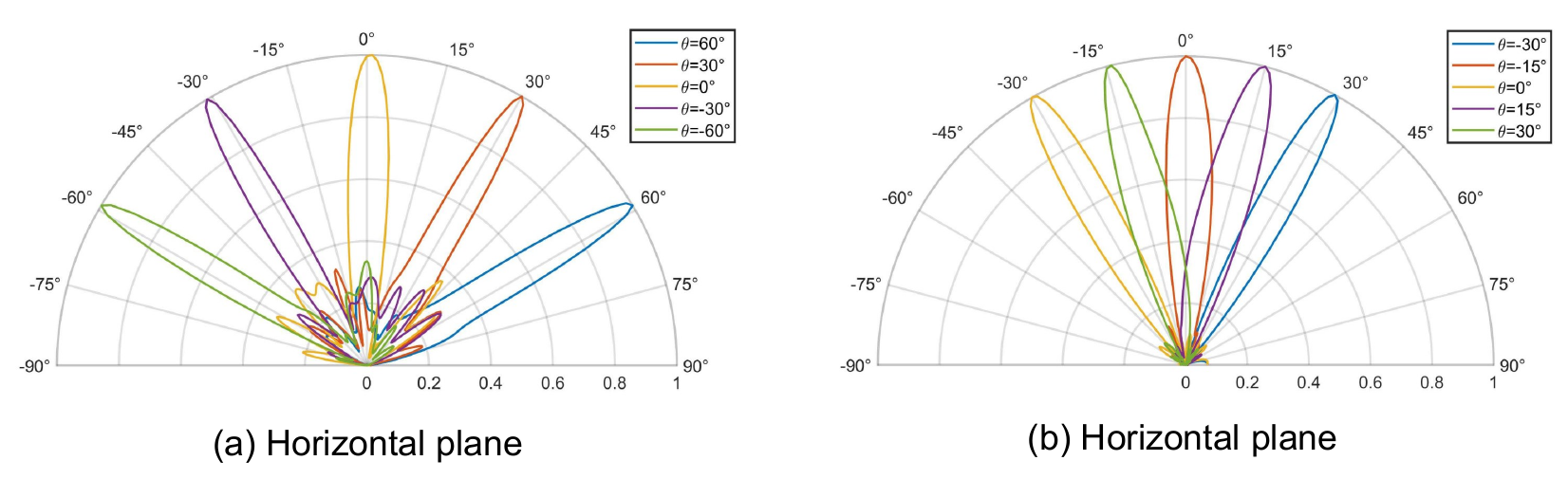}
	\caption{Normalized far-field beam patterns of the RHS: (a) horizontal plane and (b) vertical plane.}
	\label{beampattern}
\end{figure*}

Fig. \ref{beampattern} show the normalized far-field beam pattern of the 2D RHS with the desired direction from -60$^\circ$ to 60$^\circ$ in the horizontal plane and -30$^\circ$ to 30$^\circ$ in the vertical plane, respectively. This validates the capability of the 2D RHS to achieve 3D holographic beamforming. It should be noted that the aperture of the RHS could be changeable if we fixed some elements to the OFF state, which could be a special implementation of fluid antennas \cite{KAKY-2021} or movable antennas \cite{LWR-2024}.

\begin{figure*}[t]
	\centering
	\includegraphics[width=0.8\textwidth]{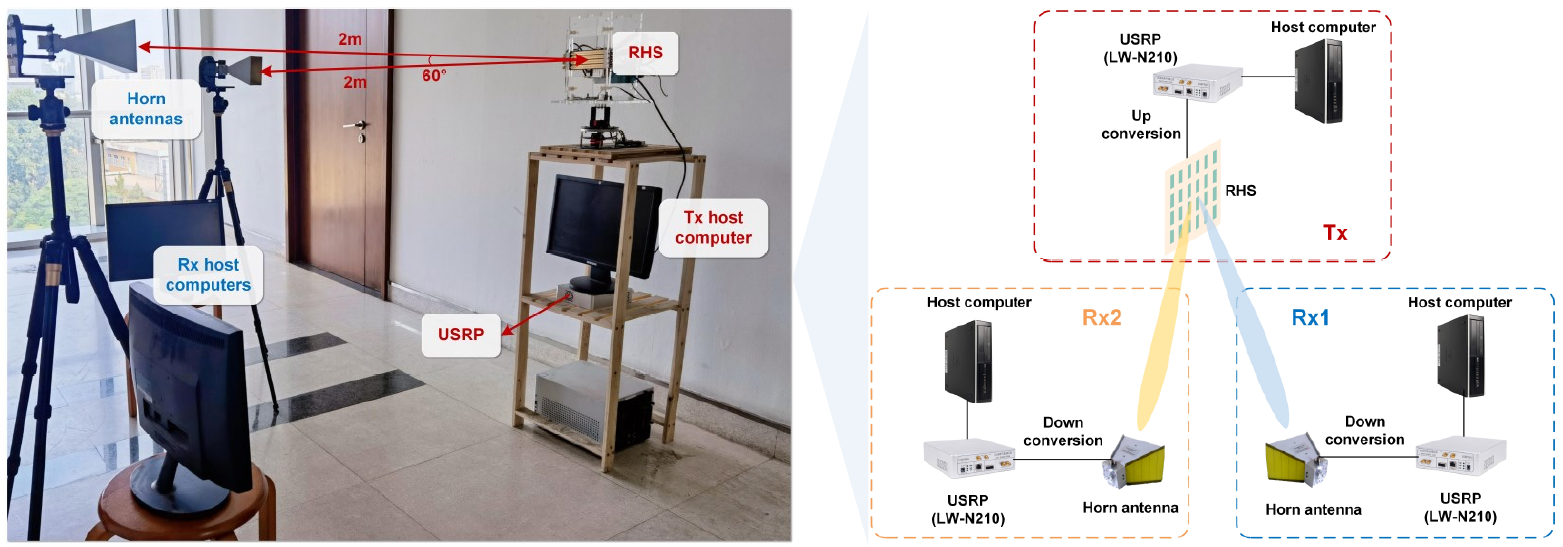}
	\caption{An RHS-aided wireless communication platform.}
	\label{platform}
\end{figure*}

\subsection{RHS-aided Wireless Communication Prototype}
\label{sec:wireless_prototype}
In this part, we will demonstrate how to develop a wireless communication platform\footnote{Please refer to \url{https://www.futurecomm-pku.cn/en/} for more details of the prototype.} based on the implemented 2D RHS. The wireless communication platform is shown in Fig.~\ref{platform}. The hardware modules of the platform are composed of one transmitter (Tx) and two receivers (Rx). The functions of these hardware modules are introduced as below \cite{RYHBHL-2023}.
\begin{itemize}
	\item \emph{Transmitter:} The Tx is implemented by a universal software radio peripheral (USRP) LW-N210. Based on a GNU radio software development kit, the Tx USRP can perform baseband signal processing and realize modulation~\cite{HSBYMMZHL-2022}. The output port of the USRP is connected to a frequency converter to up-convert the signal frequency to the working frequency of the RHS.
	
	\item \emph{Receiver:} Two Rxs are implemented. Each Rx is composed of one USRP LW-N210 and one standard horn antenna (LB-75-20-C-SF). The Rx USRP is also connected to a frequency converter to down-convert the received signal to the baseband. Then the Rx USRP then conducts demodulation and signal processing to recover the original signals.
	
	\item \emph{Host computer:} The Tx/Rx host computers control the Tx and Rxs via a software program, respectively. Specifically, a graphical interface is implemented at the Rx host computers to display the recovered signals and the corresponding parameters, such as the spectrum and constellation.
\end{itemize}

\begin{figure*}[t]
	\centering
	\includegraphics[width=0.8\textwidth]{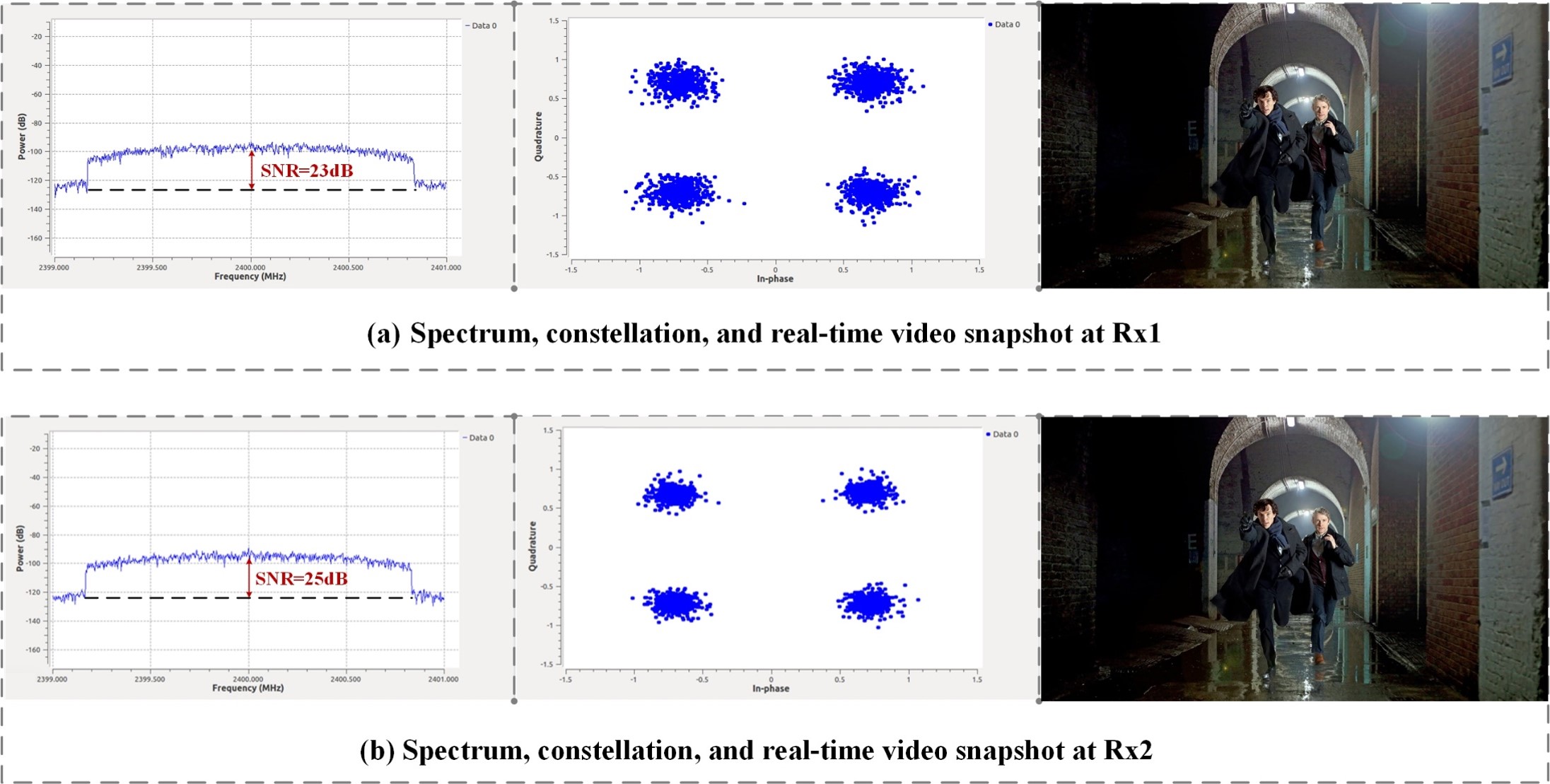}
	\caption{Measurement results of the RHS-aided wireless communication platform.}
	\label{Evaluation}
\end{figure*}

Fig.~\ref{Evaluation} depicts the graphical interface at the two Rx USRPs for the real-time video transmission using the QPSK modulation. The resolution of the video is 1920$\times$1080, and the frame rate is 30 frames/s. From the results, we can observe that the received SNR is higher than 20~dB. Moreover, the four points are equispaced around a circle, indicating a successful demodulation, and the received videos can be recovered. This validates that our RHS-aided communication platform can support real-time transmission of high-definition video.

\subsection{Holographic ISAC Prototype}
\label{sec:ISAC}
In this part, we show the prototype of an 1D-RHS-aided ISAC prototype and its measurement results \cite{HHBL-2023}.
\subsubsection{Hardware Modules}

\begin{figure}[!t]
	\centering
	\includegraphics[width=0.45\textwidth]{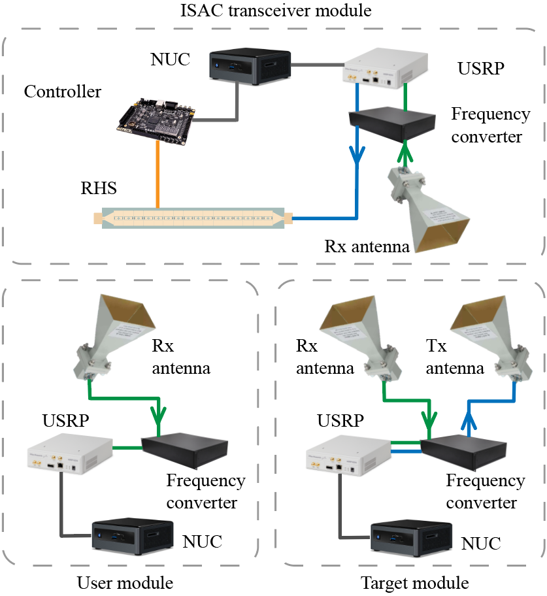}
	\caption{Illustration of the RHS-aided ISAC prototype.}
	\label{ISAC_experiment}
\end{figure}

As shown in Fig. \ref{ISAC_experiment}, the ISAC prototype consists of three modules:
\begin{itemize}

\item \textbf{Transceiver module:} This module transmits the ISAC signals using the RHS and receive echo signals from targets through a standard antenna in a time division manner. An Intel NUC connects to an FPGA to control the radiation amplitudes of the RHS as a host computer. It also connects with a USRP to simultaneously modulate the transmitted signals and demodulate the received echo signals. A frequency converter is used to up-convert and down-convert between the baseband frequency of the USRP (0-6GHz) and the working frequency of the RHS (12GHz).

\item \textbf{User module:} This module simulates a communication user, which receives and decodes the ISAC signal from the transceiver module to retrieve the communication stream. Specifically, the module first receives the ISAC signal and sends it to the frequency converter which down-converts the signal to the baseband for decoding.

\item \textbf{Target module:} This module is used to simulate radar targets by generating controllable radar echo signals~\cite{DNTYMYY-2021}. It consists of an RX antenna, an Tx antenna, a USRP, and an Intel NUC. Once the Rx antenna receives the ISAC signal transmitted by the RHS, the target module is triggered, which adds delays to the ISAC signal and emits the delayed signal through the TX antenna. The delayed time can be adjusted by a program running on the Intel NUC to simulate the targets located at different directions.
\end{itemize}

\begin{figure}[!t]
	\centering
	\includegraphics[width=0.45\textwidth]{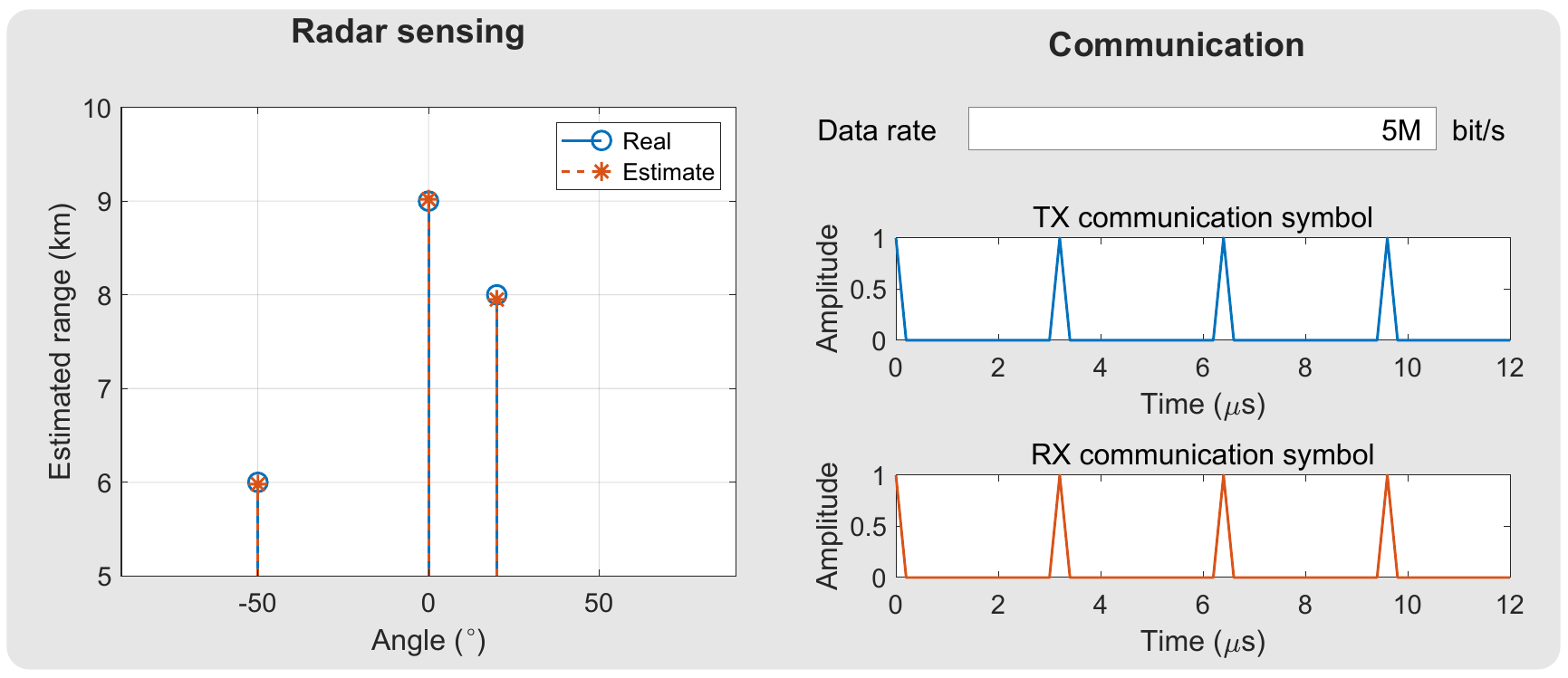}
	\caption{Radar sensing and communication performances of the holographic ISAC system.}
	\label{f_result}
\end{figure}

We evaluate the performance of the proposed ISAC system with one user and one target, which is deployed in an anechoic chamber with a size of $4$m$\times 3$m$\times 2.5$m. The user and the target modules are placed in different angle with respect to the ISAC transceiver module with an equal distance of $1.7$m. The BS first transmits the ISAC signal with a duration of $12\mu$s and then listens for the echo signal to decide the presence of the target. Fig.~\ref{f_result} shows the ISAC performance. We place the target module in directions $-50^\circ$, $0^\circ$, and $20^\circ$, respectively, to simulate the targets in different directions. The delays induced by the target module are set as $20\mu$s, $30\mu$s, and $26.7\mu$s, which corresponds to the target in $6$km, $9$km, and $8$km away. In order to sense the target, one of the main lobes of the radiation pattern is steered towards directions $-50^\circ$, $0^\circ$, and $20^\circ$ in different cycles, while the other main lobe keeps pointing toward the direction of the user module, i.e., $60^\circ$, to support the communication. It can be observed that the estimated range is close to the real range, which proves the feasibility of sensing by applying holographic ISAC. Besides, the communication symbol received by the user module is the same as the communication symbol transmitted by the BS, and the data rate between the BS and the user is $5$M bit/s, which shows that the communication between the BS and the user can be supported when the BS performs radar sensing at the same time.

\section{Outlook and Future Directions}
\label{Sec:6}
In this section, we provide a discussion on the future directions and potential challenges for the RHS-aided wireless communications. 

\subsection{Fundamental Performance Limits}
The RHS has provided a cost-effective way to realize an extremely large antenna, but also bring some new characteristics on theoretical foundations of the RHS, such as degree-of-freedom (DoF) and system capacity. In particular, one of the most prominent design shifts is from the traditional digital domain to the EM domain \cite{JMMMH-2023}, which unveil the fundamental limitations of RHS-aided communications. In addition, the radiation power of an RHS element is influenced by its parameters, such as the geometry and amplitude quantization bits, and thus it is also necessary to investigate the impact of the parameters on the performance limits.

The DoF metric indicates the number of independent data streams transmitted at the same time. The DoF of RHS-aided wireless communications has attracted the attention from the research community. For example, the work in \cite{D-2020} explored the DoF of an RHS-aided point-to-point wireless communication system. A closed-form expression of the DoF was derived, which reveals that the DoF of a line-of-Sight (LoS) channel is related to geometric factors and the wavelength, and can be higher than one. The authors in \cite{ATL-2020} considered a more general setting where a spatially-constrained aperture with a random electromagnetic isotropic channel was used. Their results revealed that the DoF is proportional to its surface area, which indicates how many antennas should be deployed in a given space such that the discrete array can achieve an approximate capacity of a continuous one.

Some papers focus on the system capacity analysis for the RHS-aided wireless networks. The authors in \cite{RBHL-2022} showed some interesting studies on the system capacity for RHS-aided wireless networks with LoS channels. Their results showed that the normalized capacity has a upper bound limit as the user density grows to infinity and the wavelength approaches zero. The study in \cite{LCGWZMC-2022} considered the non-LoS (NLoS) case and revealed that as the element spacing decreases, the mutual coupling among elements will deteriorate the system capacity. The authors in \cite{YJYYG-2023} provided the capacity analysis under the practical angle distribution and array aperture constraints, and their results showed that the capacity is significantly influenced by the angle distribution at the high SNR region, but the influence could be ignored at the low SNR region.
   
Some initial results have been reported in the literature to show the impact of some practical constraints of the RHS on the system performance. For example, the authors in \cite{XRBHL-2022} revealed the relation between discrete radiation amplitude values and the data rate in both single-user and multi-user cases, and concluded that 2 bits, i.e., 4 quantization levels of radiation amplitudes, are sufficient for the multi-user scenario while 1 bit, i.e., 2 radiation amplitudes, is already enough for the single-user case. The work in \cite{QMYL-2025} considered both the phase shift error at the RHS elements and the hardware impairment at the RF chains of the transceivers, and derived the capacity of a multi-cell RHS-aided wireless network under these practical constraints. Their results showed that these imperfect issues can be compensated by increasing the density of transmitters.

\subsection{Channel Modeling and Estimation}
With the increased antenna aperture provided by the RHS, radiating near-field region is significantly extended, where the EM waves propagation characteristics are different from that in the far-field region \cite{YZJCXR-2023}. As a result, novel channel models and channel estimation solutions are required to fit the propagation characteristics in the near-field region. On the other hand, the increased aperture means more variables in channel estimation problems, causing a prohibitively high computational complexity, which requires a low complexity algorithm \cite{EET-2016}.  

In literature, some initial attempts on channel modeling and estimation have been reported to tackle the near-field propagation characteristics brought by the RHS. For the channel modeling, the modeling introduced in \cite{ATL-2020-Model} yields a tractable and physically-meaningful spatial correlation function. Under such a function, a uniform channel models for both near and far fields can be derived. In \cite{OEL-2022}, the authors developed a channel model by considering both non-isotropic scattering and directive antennas. After the modeling, a channel estimation scheme exploiting the array geometry to identify a subspace of reduced rank that covers the eigenspace of any spatial correlation matrix was introduced as well.

For the channel estimation, the authors in \cite{SSLYB-2025} studied the hybrid near-far field channel estimation, and recognized a power diffusion effect where the power can spread when a near-field path is transformed from the spatial domain into the polar domain (or angular domain) due to a high coherence between two near-field steering vectors, which leads to an inaccurate path component estimation in the angular domain. To address this issue, the authors developed a power-diffusion aware orthogonal matching pursuit algorithm where a joint angular-polar domain to separate near-field and far-field paths. A similar idea also has been reported in \cite{YYZZ-2024}, where parameters in polar and angular domains are decomposed and constructed individually using compressed sensing methods. The authors in \cite{XYYZZ-2024} used the wavenumber domain representation, where the wavenumber is the propagation modes of EM waves if we treat the propagation channels as a waveguide. In this way, the spherical wave can be expressed as a series of plane waves \cite{ALT-2022}, and thus can be a general expression for both near-field and far-field regions and be helpful in the channel estimation.

In addition to deal with the near-field channel characteristics, existing works also aim to reduce the complexity of the channel estimation algorithms. The work in \cite{WHXSJRK-2024} proposed a self-supervised channel estimation algorithm as an efficient method to estimate the high dimensional channel matrices, without the spatial correlation relation and the ground-truth channel information. In~\cite{MR-2024}, a known correlated matrix of the channel was utilized to simplify the channel estimation process.

\subsection{Resource Management}

Subjected to the unique leakage power constraint brought by the RHS, the resource management schemes need further designs where the holographic pattern, transmit power, and channel allocations are coupled together, making the problem challenging to solve. In different application scenarios, the resource management problems need to be revisited in order to further exploit the advantage of the RHS in terms of higher spatial gain and data rates.

In literature, some works have studied the resource management issues for the RHS-aided wireless communications. The work in \cite{QMYIAL-2024} studied an energy-efficiency maximization problem where the holographic beamformer of the RHS, the digital beamformer, and the total transmit power are jointly optimized in the presence of realistic transceiver hardware impairments. The authors in \cite{AMAYSE-2023,AAYSC-2024} proposed machine learning based resource management methods to address the complex and complicated problem. The work in \cite{MIAA-2024} proposed a zero-forcing beamforming scheme for beam sweeping applications. Moreover, the author in \cite{B-2021} investigated the resource management for a wideband RHS-aided wireless communication system, where the frequency-selective nature of the RHS was considered. A joint holographic pattern and power allocation optimization scheme to alleviate the beam squint effect caused by the frequency-selective nature of the RHS, and the results show that there exists an optimal size of the RHS to strike a trade-off between the spatial diversity gain and the beam squint effect.

\subsection{Integration With Emerging Techniques}

As the RHS is an ultra-thin and lightweight antenna, which is a cost-effective solution for pursing high data rates, it can be integrated with various emerging techniques for different applications, such as sensing and localization \cite{HZYHBL-2023,TLKB-2025,JZTRJ-2023}, wireless power transfer \cite{KMUK-2023}, secure communication \cite{YJXTH-2024}, edge intelligence \cite{DTXR-2023}, and so on. In the following, we will give some examples to elaborate on how to incorporate the RHS with these emerging techniques.

\textbf{Sensing and Localization:} In vehicular networks \cite{MDJHJZSZ-2023} and industry internet of things networks~\cite{EASUM-2018}, sensing and localization capability is critical. With a larger antenna aperture provided by the RHS, it can generate a narrow beam, which can have a higher sensing and localization resolution. However, the use of the RHS for sensing and localization applications are different from that for communication as we hope that the signals at different positions should be as different as possible. Some studies have been executed for applying the RHS into sensing and localization use cases. For example, the authors in \cite{HZYHBL-2023} studied a collaborative RHS aided simultaneous localization and mapping (SLAM) scheme using wireless signals for autonomous driving applications. The work in \cite{TLKB-2025} extended the holographic ISAC system and utilized the RIS to customize the propagation environment for a better sensing and communication performance. In \cite{JZTRJ-2023}, a multi-band RHS was utilized to realize appropriate beam patterns for higher localization precision by leveraging a larger bandwidth, and a federated learning framework was proposed to provide environment-adaptive localization services.

\textbf{Wireless Power Transfer:} Wireless power transfer, also known as wireless charging, is a technology to convey energy from one device to another device through wireless signals \cite{XRC-2013}. With the increased antenna aperture, a narrower beam can be generated to transfer the power more efficiently. In addition, to fully utilize the near field region provided by the RHS can also enhance the power transfer efficiency. For example, the work in \cite{KMUK-2023} introduced an RHS-aided architecture design for wireless power transfer to achieve a higher power transfer efficiency. The authors in \cite{HNFDMY-2022} examined the possibility of wireless power transfer in the near-field region, and discussed the beam focusing scheme, hardware aspects, and its simultaneous operation with wireless communications.

\textbf{Edge Intelligence:} Edge intelligence is a set of connected systems and devices for data collection, caching, processing, and analysis in locations close to where data is captured with machine learning tools \cite{MDHJZSZ-2023}. During the procedure, it is required to transfer a large volume of data among edge servers and devices. From this viewpoint, RHS is a natural fit for edge intelligence systems which seek for lower communication delay and faster processing, especially for the device-edge information transfer \cite{SQKBHWDZH-2024}. In \cite{DTXR-2023}, the authors proposed an RHS-aided distributed edge computing scheme, where the mobile device exploits the RHS to distribute the computing tasks to multiple edge nodes simultaneously to minimize the overall latency of the system.

\section{Conclusions}
\label{Sec:7}
In this paper, we have provided a comprehensive tutorial on the RHS, which is a cost-efficient alternate to traditional phased arrays for realizing ultra-massive MIMO. We first introduced the working principles of the RHS and its unique leakage power constraint. We also provided some case studies to show how the RHS improve the performance of wireless communication and sensing. In what follows, we have demonstrated the implementation of the RHS. Based on the RHS, we have further built RHS-aided wireless communication prototype, and have reported the measurement results to verify the benefits of the RHS compared to the phased array. Finally, we have also highlighted several potential research directions for the RHS-aided wireless networks. We hope that this paper can be a useful resource for future research on RHS-aided wireless networks, utilizing its potential to provide an extremely large antenna aperture.


\begin{thebibliography}{}
\providecommand{\url}[1]{#1}
\csname url@samestyle\endcsname
\providecommand{\newblock}{\relax}
\providecommand{\bibinfo}[2]{#2}
\providecommand{\BIBentrySTDinterwordspacing}{\spaceskip=0pt\relax}
\providecommand{\BIBentryALTinterwordstretchfactor}{4}
\providecommand{\BIBentryALTinterwordspacing}{\spaceskip=\fontdimen2\font plus
\BIBentryALTinterwordstretchfactor\fontdimen3\font minus
  \fontdimen4\font\relax}
\providecommand{\BIBforeignlanguage}[2]{{%
\expandafter\ifx\csname l@#1\endcsname\relax
\typeout{** WARNING: IEEEtran.bst: No hyphenation pattern has been}%
\typeout{** loaded for the language `#1'. Using the pattern for}%
\typeout{** the default language instead.}%
\else
\language=\csname l@#1\endcsname
\fi
#2}}
\providecommand{\BIBdecl}{\relax}
\BIBdecl

\end{thebibliography}


\begin{thebibliography}{1}
\bibliographystyle{IEEEtran}

\bibitem{SOBM-2020}
S. Dang, O. Amin, B. Shihada, and M.-S. Alouini, ``What Should 6G Be?" \emph{Nature Electron.}, vol. 3, no. 1, pp. 20-29, Jan. 2020. 

\bibitem{KWYJY-2019}
K. B. Letaief, W. Chen, Y. Shi, J. Zhang, and Y.-J. A. Zhang, ``The Roadmap to 6G: AI Empowered Wireless Networks," \emph{IEEE Commun. Mag.}, vol. 57, no. 8, pp. 84–90, Aug. 2019.

\bibitem{ZJHWBD-2024}
Z. Wang, J. Zhang, H. Du, W. E. I. Sha, B. Ai, and D. Niyato, ``Extremely Large-Scale MIMO: Fundamentals, Challenges, Solutions, and Future Directions," \emph{IEEE Wireless Commun.}, vol. 31, no. 3, pp. 117-124, Jun. 2024.

\bibitem{IHAAKEI-2018}
I. Ahmed, H. Khammari, A. Shahid, A. Musa, K. S. Kim, E. D. Poorter, and I. Moerman, ``A Survey on Hybrid Beamforming Techniques in 5G: Architecture and System Model Perspectives," \emph{IEEE Commun. Surveys Tuts.}, vol. 20, no. 4, pp. 3060-3097, 4th Quart. 2018.

\bibitem{SHBYMZHL-2022}
S. Zhang, H. Zhang, B. Di, Y. Tan, M. D. Renzo, Z. Han, H. V. Poor, and L. Song, ``Intelligent Omni-Surfaces: Ubiquitous Wireless Transmission by Reflective-Refractive Metasurfaces," \emph{IEEE Trans. Wireless Commun.}, vol. 21, no. 1, pp. 219-233, Jan. 2022.

\bibitem{SJTN-2021}
S. Zhang, J. Liu, T. K. Rodrigues, and N. Kato, ``Deep Learning Techniques for Advancing 6G Communications in the Physical Layer," \emph{IEEE Wireless Commun.}, vol. 28, no. 5, pp. 141-147, Oct. 2021.

\bibitem{LWR-2024}
L. Zhu, W. Ma and R. Zhang, ``Movable Antennas for Wireless Communication: Opportunities and Challenges," \emph{IEEE Commun. Mag.}, vol. 62, no. 6, pp. 114-120, Jun. 2024.

\bibitem{XQR-2024}
X. Shao, Q. Jiang, and R Zhang, ``6D Movable Antenna based on User Distribution: Modeling and Optimization." \emph{arXiv:2403.08123}, 2024.

\bibitem{RBHYL-2022}
R. Deng, B. Di, H. Zhang, Y. Tan, and L. Song, ``Reconfigurable Holographic Surface-Enabled Multi-User Wireless Communications: Amplitude-Controlled Holographic Beamforming," \emph{IEEE Trans. Wireless Commun.}, vol. 21, no. 8, pp. 6003-6017, Aug. 2022.

\bibitem{RBHDZHL-2021}
R. Deng, B. Di, H. Zhang, D. Niyato, Z. Han, H. V. Poor, and L. Song, ``Reconfigurable Holographic Surfaces for Future Wireless Communications," \emph{IEEE Wireless Commun.}, vol. 28, no. 6, pp. 126-131, Dec. 2021.

\bibitem{MAMMCJS-2020}
M. Di Renzo, A. Zappone, M. Debbah, M.-S. Alouini, C. Yuen, J. Rosny, and S. Tretyakov, ``Smart Radio Environments Empowered by Reconfigurable Intelligent Surfaces: How It Works, State of Research, and The Road Ahead," \emph{IEEE J. Sel. Areas Commun.}, vol. 38, no. 11, pp. 2450-2525, Nov. 2020.

\bibitem{QSBCR-2021}
Q. Wu, S. Zhang, B. Zheng, C. You, and R. Zhang, ``Intelligent Reflecting Surface-aided Wireless Communications: A Tutorial," \emph{IEEE Trans. Commun.}, vol. 69, no. 5, pp. 3313-3351, May 2021.

\bibitem{YBN-2022}
Y. Zhu, B. Mao, and N. Kato, ``Intelligent Reflecting Surface in 6G Vehicular Communications: A Survey," \emph{IEEE Open J. Veh.Technol.}, vol. 3, pp. 266-277, May 2022.

\bibitem{BHLYZH-2020}
B. Di, H. Zhang, L. Song, Y. Li, Z. Han, and H. V. Poor, ``Hybrid Beamforming for Reconfigurable Intelligent Surface based Multi-User Communications: Achievable Rates With Limited Discrete Phase Shifts," \emph{IEEE J. Sel. Areas Commun.}, vol. 38, no. 8, pp. 1809-1822, Aug. 2020.

\bibitem{QR-2019}
Q. Wu and R. Zhang, ``Intelligent Reflecting Surface Enhanced Wireless Network via Joint Active and Passive Beamforming," \emph{IEEE Trans. Wireless Commun.}, vol. 18, no. 11, pp. 5394-5409, Nov. 2019.

\bibitem{XYLJ-2025}
X. He, Y. Gong, L. Huang, and J. Wang, ``Linear Complexity Holographic Beamforming For Satellite Broadcasting," \emph{IEEE Trans. Veh. Technol.}, to be published.

\bibitem{RBHHL-2022}
R. Deng, B. Di, H. Zhang, H. V. Poor, and L. Song, ``Holographic MIMO for LEO Satellite Communications Aided by Reconfigurable Holographic Surfaces," \emph{IEEE J. Sel. Areas Commun.}, vol. 40, no. 10, pp. 3071-3085, Oct. 2022.



\bibitem{XRBHL-2023}
X. Hu, R. Deng, B. Di, H. Zhang, and L. Song, ``Holographic Beamforming for LEO Satellites," \emph{IEEE Commun. Lett.}, vol. 27, no. 10, pp. 2717-2721, Oct. 2023.

\bibitem{ASDS-2013}
A. E. Forooshani, S. Bashir, D. G. Michelson, and S. Noghanian, ``A Survey of Wireless Communications and Propagation Modeling in Underground Mines," \emph{IEEE Commun. Surveys Tuts.}, vol. 15, no. 4, pp. 1524-1545, 4th Quart. 2013.


\bibitem{SG-2020}
S. Tahcfulloh and G. Hendrantoro, ``FPMIMO: A General MIMO Structure with Overlapping Subarrays for Various Radar Applications," \emph{IEEE Access}, vol. 8, pp. 11248–11267, 2020.


\bibitem{MSYLYMGEKCA-2017}
M. Xiao, S. Mumtaz, Y. Huang, L. Dai, Y. Li, M. Matthaiou, G. K. Karagiannidis, E. Bj\"ornson, K. Yang, C.-L. I, and A. Ghosh, ``Millimeter Wave Communications for Future Mobile Networks," \emph{IEEE J. Sel. Areas Commun.}, vol. 35, no. 9, pp. 1909-1935, Sept. 2017.

\bibitem{ICS-2018}
I. F. Akyildiz, C. Han and S. Nie, ``Combating the Distance Problem in the Millimeter Wave and Terahertz Frequency Bands," \emph{IEEE Commun. Mag.}, vol. 56, no. 6, pp. 102-108, Jun. 2018.

\bibitem{QBCLKXWBHEMZR-2024}
Q. Wu, B. Zheng, C. You, L. Zhu, K. Shen, X. Shao, W. Mei, B. Di, H. Zhang, E. Basar, M. D. Renzo, Z.-Q. Luo, and R. Zhang, ``Intelligent Surfaces Empowered Wireless Network: Recent Advances and the Road to 6G," \emph{Proc. IEEE}, to be published.

\bibitem{YRQJHH-2019}
Y.-C. Liang, R. Long, Q. Zhang, J. Chen, H. V. Cheng, and H. Guo, ``Large Intelligent Surface/antennas (LISA): Making Reflective Radios Smart," \emph{J. Commun. Inf. Netw.}, vol. 4, no. 2, pp. 40–50, Jun. 2019.

\bibitem{HB-2022}
H. Zhang and B. Di, ``Intelligent Omni-Surfaces: Simultaneous Refraction and Reflection for Full-Dimensional Wireless Communications," \emph{IEEE Commun. Surveys Tuts.}, vol. 24, no. 4, pp. 1997-2028, 4th-Quart. 2022.

\bibitem{MHLKZG-2020}
M. A. ElMossallamy, H. Zhang, L. Song, K. G. Seddik, Z. Han, and G. Y. Li, ``Reconfigurable Intelligent Surfaces for Wireless Communications: Principles, Challenges, and Opportunities," \emph{IEEE Trans. Cogn. Commun. Netw.}, vol. 6, no. 3, pp. 990-1002, Sep. 2020.

\bibitem{YXXTJMN-2021}
Y. Liu, X. Liu, X. Mu, T. Hou, J. Xu, M. D. Renzo, and N. A.-Dhahir, ``Reconfigurable Intelligent Surfaces: Principles and Opportunities," \emph{IEEE Commun. Surveys Tuts.}, vol. 23, no. 3, pp. 1546–1577, 3rd Quart., 2021.

\bibitem{CSGACRMM-2020}
C. Huang, S. Hu, G. C. Alexandropoulos, A. Zappone, C. Yuen, R. Zhang, M. D. Renzo, and M. Debbah, ``Holographic MIMO Surfaces for 6G Wireless Networks: Opportunities, Challenges, and Trends," \emph{IEEE Wireless Commun.}, vol. 27, no. 5, pp. 118-125, Oct. 2020.

\bibitem{TPRCGLZMHC-2024}
T. Gong, P. Gavriilidis, R. Ji, C. Huang, G. C. Alexandropoules, L. Wei, Z. Zhang, M. Debbah, H. V. Poor, and C. Yuen, ``Holographic MIMO Communications: Theoretical Foundations, Enabling Technologies, and Future Directions," \emph{IEEE Commun. Surveys Tuts.}, vol. 26, no. 1, pp. 196-257, 4th Quart. 2024.

\bibitem{P-1996}
P. Hariharan, \emph{Optical Holography: Principles, Techniques and Applications}, Cambridge University Press, 1996.

\bibitem{BJJJD-2010}
B. H. Fong, J. S. Colburn, J. J. Ottusch, J. L. Visher, and D. F. Sievenpiper, ``Scalar and Tensor Holographic Artificial Impedance Surfaces," \emph{IEEE Trans. Antennas Propag.}, vol. 58, no. 10, pp. 3212-3221, Oct. 2010.

\bibitem{OD-2017}
O. Yurduseven and D. R. Smith, ``Dual-polarization Printed Holographic Multibeam Metasurface Antenna," \emph{IEEE Antennas Wireless Propag. Lett.}, vol. 16, pp. 2738-2741, Aug. 2017.

\bibitem{MSN-2016}
M. Johnson, S. Brunton, N. Kundtz, and N. Kutz, ``Extremum-seeking Control of the Beam Pattern of A Reconfigurable Holographic Metamaterial Antenna," \emph{J. Opt. Soc. Amer. A}, vol. 33, no. 1, pp. 59-68, Jan. 2016.

\bibitem{RBHYL-2021}
R. Deng, B. Di, H. Zhang, Y. Tan and L. Song, ``Reconfigurable Holographic Surface: Holographic Beamforming for Metasurface-Aided Wireless Communications," \emph{IEEE Trans. Veh. Technol.}, vol. 70, no. 6, pp. 6255-6259, Jun. 2021.

\bibitem{FW-2016}
F. Sohrabi and W. Yu, ``Hybrid digital and analog beamforming design for large-scale antenna arrays," \emph{IEEE J. Sel. Topics Signal Process.}, vol. 10, no. 3, pp. 501-513, Apr. 2016.

\bibitem{FDBETF-2013}
F. Rusek, D. Persson, B. K. Lau, E. G. Larsson, T. L. Marzetta, and F. Tufvesson, ``Scaling Up MIMO: Opportunities and Challenges With Very Large Arrays," \emph{IEEE Signal Process. Mag.}, vol. 30, no. 1, pp. 40-60, Jan. 2013.

\bibitem{KW-2016}
K. Shen and W. Yu, ``Load and Interference Aware Joint Cell Association and User Scheduling in Uplink Cellular Networks," in \emph{Proc. IEEE SPAWC}, Edinburgh, UK, Jul. 2016.

\bibitem{SL-2004}
S. Boyd, and L. Vandenberhe, \emph{Convex Optimization}. Cambridge University Press, 2004.

\bibitem{RBHL-2022}
R. Deng, B. Di, H. Zhang and L. Song, ``HDMA: Holographic-Pattern Division Multiple Access," \emph{IEEE Journal on Selected Areas in Communications}, vol. 40, no. 4, pp. 1317-1332, April 2022.

\bibitem{DMDPA-2003}
D. Gesbert, M. Shafi, D.-S. Shiu, P. J. Smith, and A. Naguib, ``From Theory to Practice: An Overview of MIMO Space-Time Coded Wireless Systems," \emph{IEEE J. Sel. Areas Commun.}, vol. 21, no. 3, pp. 281-302, Apr. 2003.

\bibitem{Pi-2019}
``Holographic Beamforming and Phased Arrays", Pivotal Commware, Kirkland, WA, 2019.

\bibitem{HSHTH-2024}
H. Chen, S. Zeng, H. Guo, T. Svensson, and H. Zhang, ``Near-Far Field Channel Modeling for Holographic MIMO Using Expectation-Maximization Methods," in \emph{Proc. IEEE WCNC}, Dubai, UAE, Apr. 2024.

\bibitem{YZJCXR-2023}
Y. Liu, Z. Wang, J. Xu, C. Ouyang, X. Mu, and R. Schober, ``Near-Field Communications: A Tutorial Review," \emph{IEEE Open J. Commun. Soc.}, vol. 4, pp. 1999-2049, 2023.


\bibitem{YBHL-2023}
Y. Zhang, B. Di, H. Zhang, and L. Song, ``Near-Far Field Beamforming for Holographic Multiple-Input Multiple-Output," \emph{J. Commun. Inf. Netw.}, vol. 8, no. 2, pp. 99-110, Jun. 2023.

\bibitem{SBHH-2024}
S. Zhang, B. Di, H. Zhang, and H. V. Poor, ``Hierarchical Codebook Design Using Scale-Changeable Reconfigurable Holographic Surfaces in Near-Far Field Communications," in \emph{Proc. IEEE GLOBECOM}, Cape Town, South Africa, Dec. 2024.

\bibitem{HBKZHL-2022}
H. Zhang, B. Di, K. Bian, Z. Han, H. V. Poor, and L. Song, ``Toward Ubiquitous Sensing and Localization With Reconfigurable Intelligent Surfaces," \emph{Proc. IEEE}, vol. 110, no. 9, pp. 1401-1422, Sep. 2022.

\bibitem{EMOSAGMKYM-2018}
E. S. Lohan, M. Koivisto, O. Galinina, S. Andreev, A. Tolli, G. Destino, M. Costa, K. Leppanen, Y. Koucheryavy, and M. Valkama, ``Benefits of Positioning-aided Communication Technology in High-frequency Industrial IoT," \emph{IEEE Commun. Mag.}, vol. 56, no. 12, pp. 142–148, Dec. 2018.

\bibitem{HDYJYS-2017}
H. Wang, D. Zhang, Y. Wang, J. Ma, Y. Wang, and S. Li, ``RT-Fall: A Real-time and Contactless Fall Detection System with Commodity WiFi Devices," \emph{IEEE Trans. Mobile Comput.}, vol. 16, no. 2, pp. 511–526, Feb. 2017.

\bibitem{JQMY-2018}
J. Wei, Q. Gao, M. Pan, and Y. Fang, ``Device-free Wireless Sensing: Challenges, Opportunities, and Applications," \emph{IEEE Netw.}, vol. 32, no. 2, pp. 132–137, Mar. 2018.

\bibitem{JHYYC-2020}
J. Liu, H. Liu, Y. Chen, Y. Wang, and C. Wang, ``Wireless Sensing for Human Activity: A Survey," \emph{IEEE Commun. Surveys Tuts.}, vol. 22, no. 3, pp. 1629–1645, 3rd Quart., 2020.

\bibitem{KMVBS-2019}
K. V. Mishra, M. R. Bhavani Shankar, V. Koivunen, B. Ottersten, and S. A. Vorobyov, ``Toward Millimeter-Wave Joint Radar Communications: A Signal Processing Perspective," \emph{IEEE Signal Process. Mag.}, vol. 36, no. 5, pp. 100-114, Sept. 2019.

\bibitem{XHHB-2023}
X. Zhang, H. Zhang, H. Zhang, and B. Di, ``Holographic Radar: Target Detection Enabled by Reconfigurable Holographic Surfaces," \emph{IEEE Commun. Lett.}, vol. 27, no. 1, pp. 332-336, Jan. 2023.

\bibitem{XHHLB-2022}
X. Zhang, H. Zhang, H. Zhang, L. Liu, and B. Di, ``Parameter Estimation for Reconfigurable Holographic Surfaces enabled Radars," in \emph{Proc. IEEE ISWCS}, Hangzhou, China, Oct. 2022.

\bibitem{HHBZL-2023}
H. Zhang, H. Zhang, B. Di, Z. Han, and L. Song, ``Holographic Radar: Optimal Beamformer Design for Detection Accuracy Maximization," in \emph{Proc. IEEE Radarconf.}, San Antonio, Texas, May 2023.

\bibitem{P-2019}
P. Commware, \emph{Holographic Beamforming and Phased Arrays}. Washington, DC, USA: Kirkland, 2019.

\bibitem{HBKZHL-2022}
H. Zhang, B. Di, K. Bian, Z. Han, H. V. Poor, and L. Song, ``Toward Ubiquitous Sensing and Localization With Reconfigurable Intelligent Surfaces," \emph{Proc. IEEE}, vol. 110, no. 9, pp. 1401-1422, Sept. 2022.

\bibitem{XHLZHB-2025}
X. Zhang, H. Zhang, L. Liu, Z. Han, H. V. Poor, and B. Di, ``Target Detection and Positioning Aided by Reconfigurable Surfaces: Reflective or Holographic?," \emph{IEEE Trans. Wireless Commun.}, under revision.

\bibitem{HHBMZHL-2022}
H. Zhang, H. Zhang, B. Di, M. D. Renzo, Z. Han, H. V. Poor, and L. Song, ``Holographic Integrated Sensing and Communication," \emph{IEEE J. Sel. Areas Commun.}, vol. 40, no. 7, pp. 2114-2130, Jul. 2022.

\bibitem{PJY-2007}
P. Stoica, J. Li, and Y. Xie, ``On Probing Signal Design For MIMO Radar," \emph{IEEE Trans. Signal Process.}, vol. 55, no. 8, pp. 4151-4161, Aug. 2007.


\bibitem{NNNXJF-2016}
N. Lu, N. Cheng, N. Zhang, X. S. Shen, J. W. Mark, and F. Bai, ``Wi-Fi Hotspot at Signalized Intersection: Cost-Effectiveness for Vehicular Internet Access," \emph{IEEE Trans. Veh. Technol.}, vol. 65, no. 5, pp. 3506-3518, May 2016.

\bibitem{R-2017}
R. J. Mailloux, \emph{Phased Array Antenna Handbook}. Norwood, MA: Artech house, 2017.

\bibitem{P-2021}
P. Staff, \emph{Holographic Beam Forming and Phased Arrays}. Accessed: Aug. 15, 2021. [Online]. Available: https://pivotalcommware.com/wpcontent/uploads/2019/10/HBF-vs-APA-White-Paper-2019.pdf

\bibitem{TMWJTMD-2016}
T. Sleasman, M. F. Imani, W. Xu, J. Hunt, T. Driscoll, M. S. Reynolds, and D. R. Smith, ``Waveguide-Fed Tunable Metamaterial Element for Dynamic Apertures," \emph{IEEE Antennas Wireless Propag. Lett.}, vol. 15, pp.~606-609, Jul. 2016.

\bibitem{RYHBHHL-2023}
R. Deng, Y. Zhang, H. Zhang, B. Di, H. Zhang, H. V. Poor, and L. Song, ``Reconfigurable Holographic Surfaces for Ultra-Massive MIMO in 6G: Practical Design, Optimization and Implementation," \emph{IEEE J. Sel. Areas Commun.}, vol. 41, no. 8, pp. 2367-2379, Aug. 2023.

\bibitem{KAKY-2021}
K. -K. Wong, A. Shojaeifard, K. -F. Tong, and Y. Zhang, ``Fluid Antenna Systems," \emph{IEEE Trans. Wireless Commun.}, vol. 20, no. 3, pp. 1950-1962, Mar. 2021.

\bibitem{RYHBHL-2023}
R. Deng, Y. Zhang, H. Zhang, B. Di, H. Zhang, and L. Song, ``Reconfigurable Holographic Surface: A New Paradigm to Implement Holographic Radio," \emph{IEEE Veh. Technol. Mag.}, vol. 18, no. 1, pp. 20-28, Mar. 2023.

\bibitem{HHBL-2023}
H. Zhang, H. Zhang, B. Di, and L. Song, ``Holographic Integrated Sensing and Communications: Principles, Technology, and Implementation," \emph{IEEE Commun. Mag.}, vol. 61, no. 5, pp. 83-89, May 2023.

\bibitem{HSBYMMZHL-2022}
H. Zhang, S. Zeng, B. Di, Y. Tan, M. D. Renzo, M. Debbah, Z. Han, H. V. Poor, and L. Song, ``Intelligent Omni-Surfaces for Full-Dimensional Wireless Communications: Principles, Technology, and Implementation," \emph{IEEE Commun. Mag.}, vol. 60, no. 2, pp. 39-45, Feb. 2022.

\bibitem{DNTYMYY-2021}
D. Ma, N. Shlezinger, T. Huang, Y. Shavit, M. Namer, Y. Liu, and Y. C. Eldar, ``Spatial Modulation for Joint Radar-Communications Systems: Design, Analysis, and Hardware Prototype," \emph{IEEE Trans. Veh. Technol.}, vol. 70, no. 3, pp. 2283-2298, Mar. 2021.


\bibitem{JMMMH-2023}
J. C. Ruiz-Sicilia, M. D. Renzo, M. D. Migliore, M. Debbah, and H. V. Poor, ``On the Degrees of Freedom and Eigenfunctions of Line-of-Sight Holographic MIMO Communications," 2023, arXiv:2308.08009.


\bibitem{D-2020}
D. Dardari, ``Communicating with Large Intelligent Surfaces: Fundamental Limits and Models," \emph{IEEE J. Sel. Areas Commun.}, vol. 38, no. 11, pp. 2526–2537, Nov. 2020.

\bibitem{ATL-2020}
A. Pizzo, T. L. Marzetta, and L. Sanguinetti, ``Degrees of Freedom of Holographic MIMO Channels," in \emph{Proc. IEEE SPAWC}, Atlanta, GA, USA, May 2020.


\bibitem{LCGWZMC-2022}
L. Wei, C. Huang, G. C. Alexandropoulos, W. E. I. Sha, Z. Zhang, M. Debbah, and C. Yuen, ``Multi-user Holographic MIMO Surfaces: Channel Modeling and Spectral Efficiency Analysis," \emph{IEEE J. Sel. Topics Signal Process.}, vol. 16, no. 5, pp. 1112–1124, Aug. 2022.

\bibitem{YJYYG-2023}
Y. Zhang, J. Zhang, Y. Zhang, Y. Yao, and G. Liu, ``Capacity Analysis of Holographic MIMO Channels With Practical Constraints," \emph{IEEE Wireless Commun. Lett.}, vol. 12, no. 6, pp. 1101-1105, Jun. 2023.

\bibitem{XRBHL-2022}
X. Hu, R. Deng, B. Di, H. Zhang and L. Song, ``Holographic Beamforming for Ultra Massive MIMO With Limited Radiation Amplitudes: How Many Quantized Bits Do We Need?," \emph{IEEE Commun. Lett.}, vol. 26, no. 6, pp. 1403-1407, Jun. 2022.

\bibitem{QMYL-2025}
Q. Li, M. El-Hajjar, Y. Sun, and L. Hanzo, ``Performance Analysis of Reconfigurable Holographic Surfaces in the Near-Field Scenario of Cell-Free Networks Under Hardware Impairments," \emph{IEEE Trans. Wireless Commun.}, to be published.

\bibitem{EET-2016}
E. Bj\"ornson, E. G. Larsson, and T. L. Marzetta, ``Massive MIMO: Ten Myths and One Critical Question," \emph{IEEE Commun. Mag.}, vol. 54, no. 2, pp. 114-123, Feb. 2016.

\bibitem{ATL-2020-Model}
A. Pizzo, T. L. Marzetta, and L. Sanguinetti, ``Spatially-stationary Model for Holographic MIMO Small-scale Fading," \emph{IEEE J. Sel. Areas Commun.}, vol. 38, no. 9, pp. 1964–1979, Sep. 2020.

\bibitem{OEL-2022}
O. T. Demir, E. Bj\"ornson, and L. Sanguinetti, ``Channel Modeling and Channel Estimation for Holographic Massive MIMO with Planar Arrays," \emph{IEEE Wireless Commun. Lett.}, vol. 11, no. 5, pp. 997–1001, May 2022.

\bibitem{SSLYB-2025}
S. Yue, S. Zeng, L. Liu, Y. C. Eldar, and B. Di, ``Hybrid Near-Far Field Channel Estimation for Holographic MIMO Communications," \emph{IEEE Trans. Wireless Commun.}, to be published.

\bibitem{YYZZ-2024}
Y. Chen, Y. Wang, Z. Wang, and Z. Han, ``Angular-Distance Based Channel Estimation for Holographic MIMO," \emph{IEEE J. Sel. Areas Commun.}, vol. 42, no. 6, pp. 1684-1702, Jun. 2024.

\bibitem{XYYZZ-2024}
X. Guo, Y. Chen, Y. Wang, Z. Wang, and Z. Han, ``Wavenumber Domain Sparse Channel Estimation in Holographic MIMO," in \emph{Proc. IEEE ICC}, Denver, CO, USA, Jun., 2024.

\bibitem{ALT-2022}
A. Pizzo, L. Sanguinetti, and T. L. Marzetta, ``Fourier Plane-wave Series Expansion for Holographic MIMO Communications," \emph{IEEE Trans. Wireless Commun.}, vol. 21, no. 9, pp. 6890–6905, Sep. 2022.

\bibitem{WHXSJRK-2024}
W. Yu, H. He, X. Yu, S. Song, J. Zhang, R. D. Murch, and K. B. Letaief, ``Learning Bayes-Optimal Channel Estimation for Holographic MIMO in Unknown EM Environments," in \emph{Proc. IEEE ICC}, Denver, CO, USA, Jun., 2024.

\bibitem{MR-2024}
M. Rezvani and R. Adve, "Channel Estimation for Dynamic Metasurface Antennas," \emph{IEEE Trans. Wireless Commun.}, vol. 23, no. 6, pp. 5832-5846, Jun. 2024.

\bibitem{QMYIAL-2024}
Q. Li, M. El-Hajjar, Y. Sun, I. Hemadeh, A. Shojaeifard, and L. Hanzo, ``Energy-Efficient Reconfigurable Holographic Surfaces Operating in the Presence of Realistic Hardware Impairments," \emph{IEEE Trans. Commun.}, vol. 72, no. 8, pp. 5226-5238, Aug. 2024.

\bibitem{AMAYSE-2023}
A. Adhikary, M. S. Munir, A. D. Raha, Y. Qiao, S. H. Hong, and E.-N. Huh, ``An Artificial Intelligence Framework for Holographic Beamforming: Coexistence of Holographic MIMO and Intelligent Omni-Surface," in \emph{Proc. ICOIN}, Bangkok, Thailand, Jan. 2023.

\bibitem{AAYSC-2024}
A. Adhikary, A. D. Raha, Y. Qiao, S. W. Kang and C. S. Hong, ``Transfer Learning Empowered Power Allocation in Holographic MIMO-enabled Wireless Network," in \emph{Proc. IEEE NOMS} Seoul, Republic of Korea, May 2024.

\bibitem{MIAA-2024}
M. Yaser Yagan, I. Hokelek, A. E. Pusane, and A. Gorçin, ``Zero-Forcing Beamforming for Beam Sweeping with Reconfigurable Holographic Surfaces," in \emph{Proc. IEEE WCNC}, Dubai, United Arab Emirates, Apr. 2024.

\bibitem{B-2021}
B. Di, "Reconfigurable Holographic Metasurface Aided Wideband OFDM Communications Against Beam Squint," \emph{IEEE Trans. Veh. Technol.}, vol. 70, no. 5, pp. 5099-5103, May 2021.


\bibitem{MDJHJZSZ-2023}
M. Xu, D. Niyato, J. Chen, H. Zhang, J. Kang, Z. Xiong, S. Mao, and Z. Han, "Generative AI-Empowered Simulation for Autonomous Driving in Vehicular Mixed Reality Metaverses," \emph{IEEE J. Sel. Topics Signal Process.}, vol. 17, no. 5, pp. 1064-1079, Sep. 2023.

\bibitem{EASUM-2018}
E. Sisinni, A. Saifullah, S. Han, U. Jennehag, and M. Gidlund, ``Industrial Internet of Things: Challenges, Opportunities, and Directions," \emph{IEEE Trans. Ind. Inform.}, vol. 14, no. 11, pp. 4724-4734, Nov. 2018.

\bibitem{HZYHBL-2023}
H. Zhang, Z. Yang, Y. Tian, H. Zhang, B. Di, and L. Song, ``Reconfigurable Holographic Surface Aided Collaborative Wireless SLAM Using Federated Learning for Autonomous Driving," \emph{IEEE Trans. Intell. Veh.}, vol. 8, no. 8, pp. 4031-4046, Aug. 2023.

\bibitem{TLKB-2025}
T. Wei, L. Wu, K. V. Mishra, and B. Shankar, ``RIS-Aided Wideband Holographic DFRC," \emph{IEEE Trans. Aerosp. Electron. Syst.}, to be published.

\bibitem{JZTRJ-2023}
J. Hu, Z. Chen, T. Zheng, R. Schober, and J. Luo, ``HoloFed: Environment-Adaptive Positioning via Multi-Band Reconfigurable Holographic Surfaces and Federated Learning," \emph{IEEE J. Sel. Areas Commun.}, vol. 41, no. 12, pp. 3736-3751, Dec. 2023.

\bibitem{XRC-2013}
X. Zhou, R. Zhang, and C. K. Ho, ``Wireless Information and Power Transfer: Architecture Design and Rate-Energy Tradeoff," \emph{IEEE Trans. Commun.}, vol. 61, no. 11, pp. 4754-4767, Nov. 2013.

\bibitem{KMUK-2023}
K. Li, M. Y. Naderi, U. Muncuk, and K. R. Chowdhury, ``MetaResonance - A Reconfigurable Surface for Holographic Wireless Power Transfer," \emph{IEEE Trans. Ind. Electron.}, vol. 70, no. 5, pp. 4682-4692, May 2023.

\bibitem{HNFDMY-2022}
H. Zhang, N. Shlezinger, F. Guidi, D. Dardari, M. F. Imani, and Y. C. Eldar, ``Near-Field Wireless Power Transfer for 6G Internet of Everything Mobile Networks: Opportunities and Challenges," \emph{IEEE Commun. Mag.}, vol. 60, no. 3, pp. 12-18, Mar. 2022.

\bibitem{YJXTH-2024}
Y. Xu, J. Liu, X. Wu, T. Guo, and H. Peng, ``Reconfigurable Holographic Surface-Assisted Wireless Secrecy Communication System," \emph{Electron.}. vol. 13, no. 7, art. 1359, pp. 1-14, Apr. 2024.

\bibitem{MDHJZSZ-2023}
M. Xu, D. Niyato, H. Zhang, J. Kang, Z. Xiong, S. Mao, and Z. Han, ``Sparks of Generative Pretrained Transformers in Edge Intelligence for the Metaverse: Caching and Inference for Mobile Artificial Intelligence-Generated Content Services," \emph{IEEE Veh. Technol. Mag.}, vol. 18, no. 4, pp. 35-44, Dec. 2023.

\bibitem{SQKBHWDZH-2024}
S. Zhang, Q. Liu, K. Chen, B. Di, H. Zhang, W. Yang, D. Niyato, Z. Han, and H. V. Poor, "Large Models for Aerial Edges: An Edge-Cloud Model Evolution and Communication Paradigm," \emph{IEEE J. Sel. Areas Commun.}, to be published.

\bibitem{DTXR-2023}
D. Geng, T. Li, X. He, and R. Jin, ``Reconfigurable Holographic Surface-Aided Distributed Edge Computing," in \emph{Proc. WCSP}, Hangzhou, China, Nov. 2023.


\end{thebibliography}
\end{document}